\documentclass[12pt,a4]{article}

\usepackage[T1]{fontenc}
\usepackage{graphicx}
\usepackage[dvips]{epsfig}
\usepackage{amssymb,amsfonts,amsmath}
\usepackage{tabularx}
\usepackage{enumerate}
\usepackage{ulem}

\begin{document}

\centerline{\Large Deep ocean early warning signals of an Atlantic MOC collapse}  
\bigskip
\centerline{ Qing Yi Feng, Jan P. Viebahn and  Henk A. Dijkstra}
\bigskip
\centerline {\small Institute for Marine and Atmospheric research Utrecht (IMAU),}
\centerline {\small Department of Physics and Astronomy, Utrecht University, Utrecht, The Netherlands.}
\bigskip

{\bf 
The Atlantic Meridional Overturning Circulation (MOC) is a crucial part of 
the  climate system because of its  associated northward heat transport 
\cite{Ganachaud2000, Johns2011}. The present-day MOC  is  sensitive 
to freshwater anomalies  and may collapse to a state with a strongly 
reduced northward heat transport \cite{Bryan1986, Rahmstorf2000}.   
A future collapse of  the Atlantic  MOC has  been identified as 
one of the  most dangerous  tipping points in  the climate system 
\cite{Lenton2008}.   It is  therefore crucial  to develop early warning  
indicators  for such a  potential collapse based on   relatively short  
time series. So far, attempts  to use indicators based on critical slowdown 
have been marginally successful  \cite{Lenton2011}. Based on complex 
climate network reconstruction \cite{Tsonis:2006tk, Donges2009},   we  
here present  a promising  new   indicator for the  MOC  collapse  that   
efficiently monitors  spatial changes  in deep ocean circulation. Through 
our analysis  of the performance of this indicator we formulate  optimal 
locations of  measurement  of the MOC to  provide early warning signals 
of a collapse.  Our results imply that an increase in spatial resolution 
of the Atlantic MOC observations  (i.e., at more sections) can improve 
early detection, because the spatial coherence in the deep ocean arising 
near the transition is better captured.  
}

To develop this indicator and to study its performance, we  use the results from 
the control simulation and the freshwater perturbed (from  now on referred to as `hosing') 
simulation of the  FAMOUS climate model (see Methods). In the hosing simulation,  
the freshwater flux over the extratropical North Atlantic is increased  linearly  from 
zero to 1.0 Sv (1 Sv = 10$^6$ m$^3$s$^{-1}$)   over 2000 years \cite{Hawkins2011}. 
Fig.~1\textbf {a}  displays the  annual mean time series of the Atlantic MOC value at latitude  
26$^\circ$N  and at 1000 m depth   for the control simulation (green curve) and 
the   hosing simulation  (blue curve).  Whereas the control MOC values are statistically  
stationary  over the 2000 year integration  period, the MOC values for the hosing 
simulation  show a  rapid decrease between the years 800-1050.  

We also use results from the FAMOUS simulations in which the freshwater flux  was fixed 
after a certain integration time and the model was integrated to equilibrium 
\cite{Hawkins2011}. Time  series of the MOC of the last 100 years of these 
simulations were analyzed and  the mean MOC values (again at  26$^\circ$N 
and 1000 m depth) are plotted  as the red dots labeled from 1 to 6 in 
Fig.~1\textbf {a}.    Figure~1\textbf {b-d} show mean  MOC streamfunction patterns 
of  three of these equilibrium simulations (labelled with 1, 4 and 6 in Fig.~1\textbf {a}). 
Although the equilibrium value of the MOC decreases with larger freshwater inflow,  
the changes in the MOC pattern are relatively minor.  

For a slightly higher value of  the freshwater flux as at point 6, the equilibrium solution 
is already a collapsed state \cite{Hawkins2011}. In the hosing simulation, the freshwater 
is, however,  added relatively fast compared to 
the  equilibration time scale  of the MOC. Hence the MOC maintains its pattern 
for much higher freshwater inflow and collapses near year 900 (Fig.~1\textbf {a}). 
To determine a collapse time $\tau_c$ more precisely, we use the control 
simulation of FAMOUS (see Methods). We find $\tau_c=874$ years 
and $\tau_c$ is shown as the red dashed line in Fig.~1\textbf {a}.

Critical  slowdown has been a key phrase in the detection of  tipping points in 
ecosystems \cite{Dakos2011} and the climate system \cite{Lenton2011}. Indeed 
in many natural and  man-made systems, there is an increase in response times to 
perturbations as a  tipping point is approached  \cite{Scheffer2009}.  The 
detection of MOC  collapses and the design of early warning  signals has so far 
mostly been  based on the analysis of single time series \cite{Livina2007, 
Lenton2011}.  Critical  slowdown induces changes in variance and lag-1 
autocorrelation  in the time  series that can be connected to the distance 
to the tipping  point and hence these quantities can serve as early 
warning indicators \cite{Scheffer2009}.  

Here we build on an idea to use complex network theory to construct an 
early warning indicator \cite{Mheen2013}. The complex network 
we employ in this study is the Pearson Correlation Climate 
Network \cite{Feng2014} abbreviated below as PCCN (see Methods 
for details on the network reconstruction).   
We first construct PCCNs using the complete Atlantic MOC field for each 
of the six 100 year equilibrium simulations (red dots in Fig.~1\textbf {a}).  
For the topological   analysis of the PCCNs,  we only  use the  degree  field  
\cite{Tsonis:2006tk,   Donges2009, Viebahn2014}.   The degree  of a node 
is the number  of links  between this  node and  other nodes.  As shown in  
Fig.~2  the changes in the degree field  at the equilibrium solutions 1, 2, 4 
and 6 in Fig.~1\textbf {a} are distinct.  When the freshwater forcing is increased, 
high degree in the network -- indicating high spatial  MOC correlations --  first appears 
at nodes in the South Atlantic at about 1000 m depth (Fig.~2\textbf {a-b}).  It
subsequently  extends to the whole  Atlantic with highest amplitudes in the deep 
ocean at  midlatitudes (Fig.~2\textbf {c-d}). 

A similar result  was obtained  using networks from the temperature field of 
an idealized spatially  two-dimensional model of the MOC \cite{Mheen2013}. 
The behavior of the degree field can be understood from the underlying structure 
of the Empirical Orthogonal Functions (EOFs) of the MOC  field (see Supplementary 
Information).  Once the transition is approached,  one of the EOFs becomes most dominant in the 
variability.  The network can be seen  as a coarse-graining of the variability and by 
focussing only on the largest correlations,  it is ideally suited to monitor the changes 
in spatial correlations of the system once the  transition is approached. 

Next, we study the transient behavior of the hosing simulation by constructing 
similar PCCNs (see Methods for details on the sliding window and threshold values 
$\tau$ used).   In \cite{Mheen2013}, the kurtosis $K_d$ of the degree distribution 
was introduced as an effective indicator to capture the changes in the topology 
of the degree field.   For the complete Atlantic MOC field,  the 
values of  $K_d$ for the hosing simulation (blue curve)  and for the control 
simulation (green curve) are plotted  in Fig.~3\textbf {a}. For the hosing simulation, 
there is indeed a  strong increase of $K_d$ to values far extending those for the
control  simulation significantly  before  the collapse time $\tau_c$.  For comparison, 
the critical slowdown indicators variance  (Fig.~3\textbf {b}) and lag-1 autocorrelation 
(Fig.~3\textbf {c})  based on the complete Atlantic MOC data (and using the same sliding window) do not 
show any early warning signal of the MOC transition before the  collapse  time  
$\tau_c$. 

To provide a measure of the performance of an indicator $I$, 
we introduce an evaluation scheme (see Methods), which consists 
of the detection time, the reliability of the indication, and the intensity 
of the indication, $\gamma^I$. The excellent performance of the kurtosis 
indicator  ${K_d}$ (with a detection time at 738 years, no false alarm and a 
positive value of $\gamma^{K_d}$) in Fig.~3\textbf {a} is shown in 
entry No.1 in Table~\ref{t:T1}. 

So far, we used the complete Atlantic MOC field of the FAMOUS model data,
but at the moment, MOC observations are only routinely made at 
26$^\circ$N through  the RAPID-MOCHA program \cite{Cunningham2007}.  
There are also initiatives  to  monitor the MOC at 35$^\circ$S (SAMOC, 
see \sloppy{www.aoml.noaa.gov/phod/SAMOC\_international}) and 
at  about 60$^\circ$N  (OSNAP, see \sloppy{www.o-snap.org}). Motivated 
by the fact that current and near future available observations  of the 
MOC will only be available along zonal sections in the Atlantic, we next 
reconstructed networks (and the indicator  $K_d$) from limited section 
MOC data of the FAMOUS model results. 

The performance of the kurtosis indicator ${K_d}$ for different 
sections and combinations of sections is shown in Table~\ref{t:T1}. 
The indicator $K_d$ provides  two false alarms (see Methods) and one 
relatively late  alarm for data at  26$^\circ$N (No. 2  in Table~\ref{t:T1}),  no 
alarm at all for data at 33$^\circ$S (No. 3  in Table~\ref{t:T1}) and does 
not perform well  for all other single section data (see Supplementary Table~1). 
When the lag-1 autocorrelation and variance 
of time series averaged over a single section are considered (see Supplementary 
Tables 2 and 3), the lag-1 autocorrelation performs best 
at 21$^\circ$N although it still gives a false alarm. The variance indicator does 
not give any warning for all single sections. 

To determine the optimal observation locations of the MOC  using $K_d$,  
we systematically deleted    sections from the complete Atlantic MOC field
and evaluated the performance of  the indicator $K_d$ for the remaining 
sections. Entry No.  4  in Table~\ref{t:T1} shows 
that the indicator $K_d$ still works well for the  
Atlantic MOC field with halved horizontal resolution (21(depth) $\times$ 21(latitude)$=441$ 
nodes),  as well as for the sets of sections including midlatitudes both in the Northern  
and Southern  Hemisphere (No. 6-8 in Table ~\ref{t:T1}). However, 
detection fails for the set of sections located only in the Northern Hemisphere at 
mid- and high latitudes (No.  5 in Table~\ref{t:T1}).   

For the latitude sets I and II, consisting of 18$^\circ$S, 13$^\circ$S, 
8$^\circ$S, 11$^\circ$N,  16$^\circ$N, 21$^\circ$N, 26$^\circ$N and 
31$^\circ$N   (blue curve  in Fig.~4 and No. 9 in Table~\ref{t:T1}) 
or  33$^\circ$S, 18$^\circ$S,  13$^\circ$S, 11$^\circ$N, 16$^\circ$N, 
21$^\circ$N, 26$^\circ$N and  31$^\circ$N (green curve in Fig.~4  
and No. 10 in Table~\ref{t:T1}),  respectively, the  performance of the indicator $K_d$ 
is comparable to the case of the  complete  Atlantic MOC field  (Fig.~3\textbf {a} 
and No.  1 in Table~\ref{t:T1}).  Both sets consist of   21(depth)$\times$8(latitude)$
=168$ grid points and hence  are considerably reduced (by 81\%) 
in comparison to  the complete  Atlantic  MOC field.  Further analysis (see Supplementary 
Table 4) indicates that the data at the  sections  18$^\circ$S and 31$^\circ$N are 
essential for a good  detection of the MOC collapse. The results of No.  
9 and 10 in Table ~\ref{t:T1} also indicate that including the MOC data at  33$^\circ$S  
(near the  SAMOC  section) would  advance  the detection time of a future collapse of the 
MOC compared to including a section at 8$^\circ$S, but the detection intensity $\gamma^{K_d}$ 
of the kurtosis indicator ${K_d}$ 
slightly decreases. 

The physical reason for the optimal observation regions is the following. In 
ocean-climate models, the MOC collapse is due to a   robust feedback 
involving the transport of salinity by the ocean circulation, the salt-advection 
feedback \cite{Stommel1961, Walin1985}. The 
subtropics are strong evaporation regions and hence the largest salinity 
gradients, central in the salt-advection feedback, 
are located in these  regions. Consequently, it is expected that here the strongest 
response due to  the salt-advection feedback appears and hence the strongest 
spatial  correlations in the MOC field. Support for this  comes
from the structure of the dominant EOFs of the equilibrium solutions  where 
the largest amplitude  is also located in these midlatitude regions (see 
Supplementary Material).  
  
Using techniques of complex network theory we have provided a
novel  indicator which  can give an early warning signal for a collapse of the 
MOC. When applied to data from the FAMOUS model \cite{Hawkins2011}, our 
results  show that  when the appropriate midlatitude North and South Atlantic   
MOC  data are  available, the kurtosis indicator  $K_d$  provides a 
strong anomalous signal at least 100 years before the transition.  
Although at the moment a collapse of the MOC is considered a high-impact 
but low-probability event \cite{IPCC2013}, modesty is required  regarding our 
confidence in this statement. The processes controlling the behavior of the MOC, 
such as the  downwelling in the northern North Atlantic boundary currents, 
are not well  represented in many of the climate models on which this 
statement is based.   In  state-of-the-art GCMs, such as
those used in the IPCC AR5,  MOC collapses  have  not been found yet 
but it is fair to say that these models have  not  been extensively tested 
for this behavior \cite{IPCC2013, Huisman2010}.  

Up to  now, there is  about 10 years  of data on the MOC from  the RAPID-MOCHA 
array, and although this array will likely be operational  up to the year 2020,  the  temporal 
extend of the  time series proceeds  only slowly.  Due to this insufficient length of the observational record, 
the indicator $K_d$ is not yet applicable to observational data and hence cannot be used to determine 
whether the recent downward trend in the observed MOC strength at  26$^\circ$N 
\cite{Smeed2013} is the start of a collapse.  Such MOC behavior 
may just  be  related to natural   variability such as that associated  with the 
Atlantic Multidecadal Oscillation \cite{Kerr2000, Knight2005}. The strong 
element of an indicator such as $K_d$ is, however,  that it  is based in spatial 
correlations.  Our results  indicate that an increase in spatial resolution 
of the MOC observations  (i.e., at more sections) can improve early detection using
this   indicator, because the spatial coherence arising near the transition is 
better captured.  This is another reason that such observations should be given 
high  priority in climate research. 

\clearpage 
\section*{Methods}
\begin{itemize} 
\item[]  {\bf Model}: The FAMOUS model is a version of the HadCM3 model with a lower 
resolution  ocean and atmosphere component; the  details of the model are 
described in \cite{Hawkins2011}.  The model has  horizontal resolution of 
2.5$^\circ$ $\times$ 3.75$^\circ$ and  a vertical resolution of 20 levels. 
Annual mean data of the  MOC field were analyzed for each simulation.

\item[] {\bf Collapse Time $\tau_c$}: With a Student t-test we first determine the time, say 
$\tau_s$, at  which  the trend of the MOC time series  of the hosing   simulation 
deviates significantly (p = 0.05)  from that of the control  simulation. Subsequently,  
the  standard deviation  $\sigma$  of the MOC values of the control simulation 
over the first $\tau_s$  years is computed.  Next, the first time when the MOC anomaly 
value (with respect to the mean of the  first $\tau_s$  years)  in the hosing simulation 
decreases below - 3$\sigma$, we consider  the MOC to start collapsing which 
defines $\tau_c$.  In the FAMOUS model, when considering the MOC values at 26$^\circ$N 
and  1000 m depth, we  find $\tau_s =$ 350 years and $\tau_c =$ 874 years. 

\item[] {\bf Network Reconstruction}:  From the MOC data of the 
FAMOUS model over the latitudinal domain [35$^\circ$S-70$^\circ$N] of the Atlantic Ocean, 
we construct a  Climate Network (CN).  The nodes of the network  are 
the latitude-depth values of the grid points of the model.   A `link'  between 
two nodes is determined by  a significant interdependence between their  
MOC anomaly time series. There  are several measures  of quantifying the 
degree of statistical interdependence,  and  the most common one is  
calculating the Pearson correlations of pairs of  time series \cite{Tsonis2004}. 
More precisely, in a Pearson Correlation Climate Network (PCCN) \cite{Feng2014} 
an unweighted and undirected link between two nodes exists if the linear Pearson 
correlation coefficient of the MOC time series at these two nodes exceeds a 
threshold value such that the correlation is significant (p=0.05).

For example, in case of  the complete Atlantic MOC field and the  100 year equilibrium 
simulations,  each network consists of 21(depth) $\times$ 42 (latitude)$=882$ nodes,
and a threshold value of $\tau=0.5$ guarantees that a Pearson correlation between MOC 
time series at two nodes exceeding $\tau$ is significant. 
For the hosing simulation, as shown in Fig.~1\textbf {a}, the 
MOC values  display a significant trend  which would 
affect the accuracy of early warning  signal detection. Hence, for the study of 
the transient behaviour of the hosing simulation,  we first linearly detrend the 
MOC field before reconstruction  of the PCCNs. Next,  a 100-year sliding window 
size (with a shift of one year)  is used and values of $\tau$  between 0.5 and 0.9 are chosen 
for each PCCN, which  guarantees significant  correlations (p = 0.05) 
in defining the links  among the nodes  in each PCCN. 

\item[] {\bf Evaluation Scheme for the Performance of an Indicator}: \\
1) {\it Detection time}: When a value of an indicator for the hosing simulation exceeds the maximum of the
same indicator for the control simulation, we flag an alarm of a collapse. For example, 
the maximum of $K_d$ for the control simulation (green curve in  Fig.~3\textbf {a}) 
is plotted as the blue dashed line in Fig.~3\textbf {a}. In this case,  an early warning 
signal is detected at year 738 which is 136 years before the MOC  collapse time  
$\tau_c= 874$ years. \\
2) {\it False Alarm}: We label a detection time which is smaller than $\tau_c - 300$ 
years  as a false alarm. The 300 years is based
on the horizon considered in IPCC-AR5 \cite{IPCC2013} for  the 
evolution of the climate system (such as reflected in the Extended 
Concentration Pathway Emissions and Forcing scenarios, ECPs). Such a time
period is also considered to be a political time horizon \cite{Lenton2008}
such that decisions taken within this period are able to affect the occurrence 
of a collapse. In addition, when the detection time is within [$\tau_c - 300, \tau_c$], 
but the signal only lasts less than 5 years it is also considered as a false alarm. \\
3) {\it Detection intensity} $\gamma^I$: The amplitude of the peak of the indicator is also 
important.  Therefore, we additionally define another quality measure $\gamma^I$ 
for indicator $I$ as
\[
\gamma^I =\frac{\max_P(I^{hosing})}{\max_{[0,\tau_c]} (I^{control})}-1
\]
Here the maximum of the control is taken over the total time 
interval $[0,\tau_c]$ and that of the hosing only over the peak interval $P$
where the indicator extends over the threshold value. The larger the value of $\gamma^I$ the better the quality of the 
indicator $I$; if $\gamma^I \leq 0$ no detection occurs. For 
example,  based  on the results in  Fig.~3\textbf {a} the  value of 
$\gamma^{K_d}$ for the complete Atlantic MOC field   is 
about 0.07 (see No. 1 in Table~\ref{t:T1}). Values of $\gamma$ 
for both the variance (Fig.~3\textbf {b}) and the lag-1 autocorrelation 
(Fig.~3\textbf {c}) are smaller than zero. 
\end{itemize} 

\section*{Correspondence}
Correspondence and requests for materials should be addressed to Q. Y. Feng
(email: Q.Feng@uu.nl). 

\section*{Author contribution}
All three authors designed the study, Q. Y. Feng performed most of the analysis, 
and all authors contributed to the writing of the paper. 

\section*{Acknowledgments}
We thank Ed Hawkins and Robin Smith  for providing the FAMOUS data and 
Jonathan Donges, Norbert Marwan and Reik Donner (PIK, Potsdam)  
for providing the software  package "pyUnicorn" used in the network  calculations here.  
The authors would like to acknowledge the support of the LINC project (no. 289447) 
funded by  the Marie-Curie ITN program (FP7-PEOPLE-2011-ITN) of EC. 
QYF also thanks all the LINC members for constructive suggestions 
and technical support.

\clearpage 

\begin{figure}[t]
\begin{center}
	\includegraphics[width=0.4\textwidth]{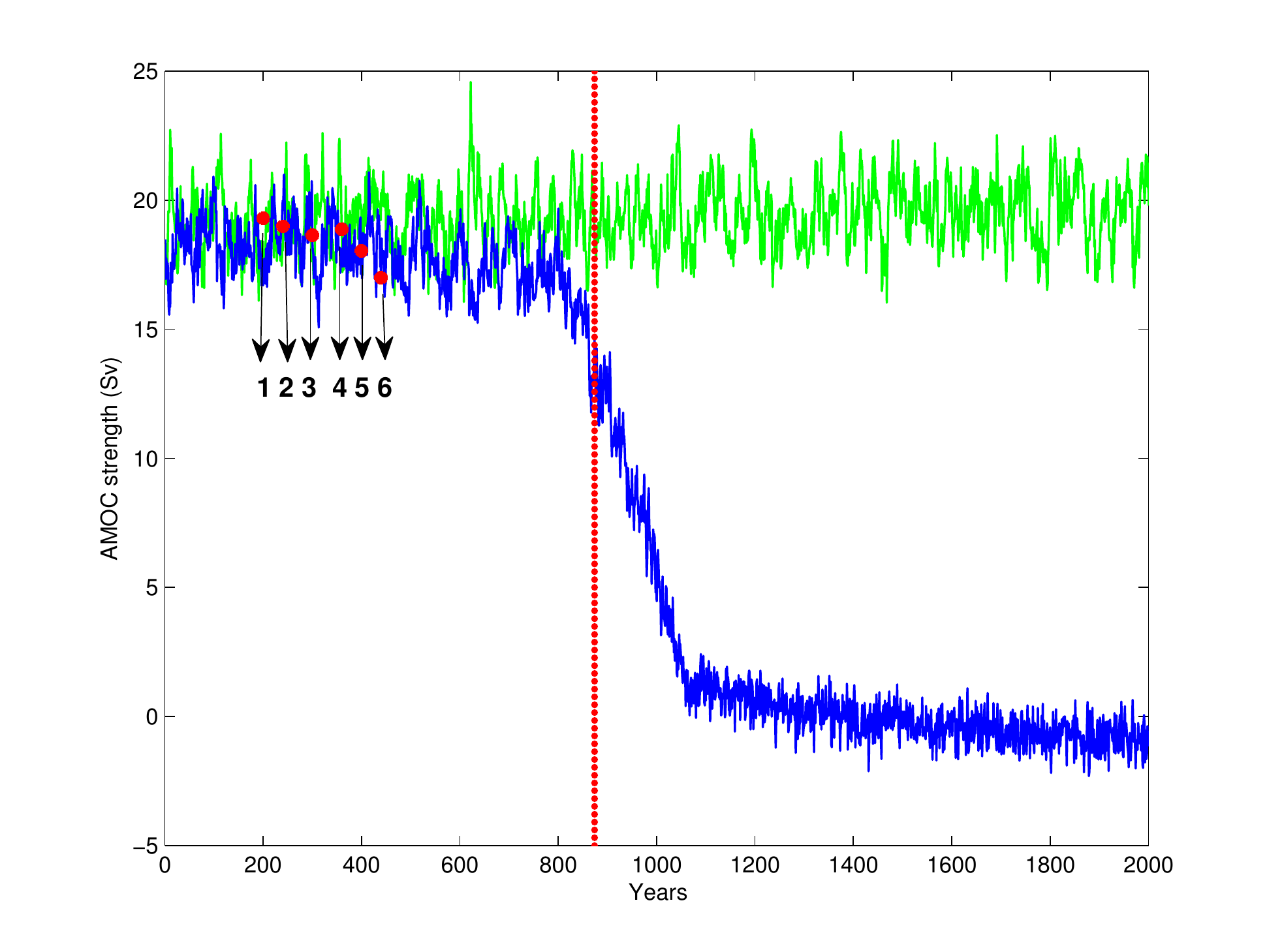} \textbf {(a)}
	\includegraphics[width=0.4\textwidth]{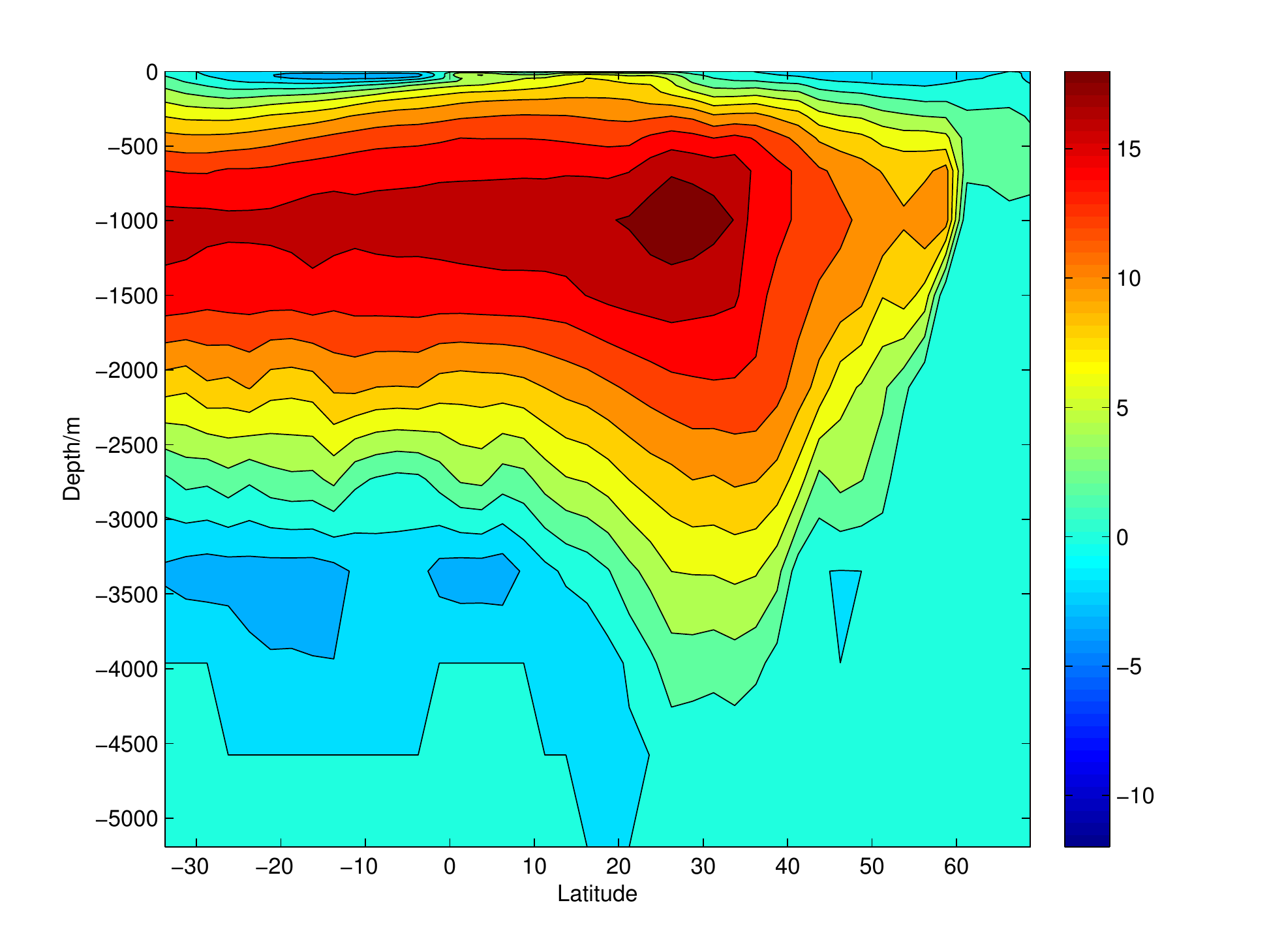} \textbf {(b)}  \\ 
	\includegraphics[width=0.4\textwidth]{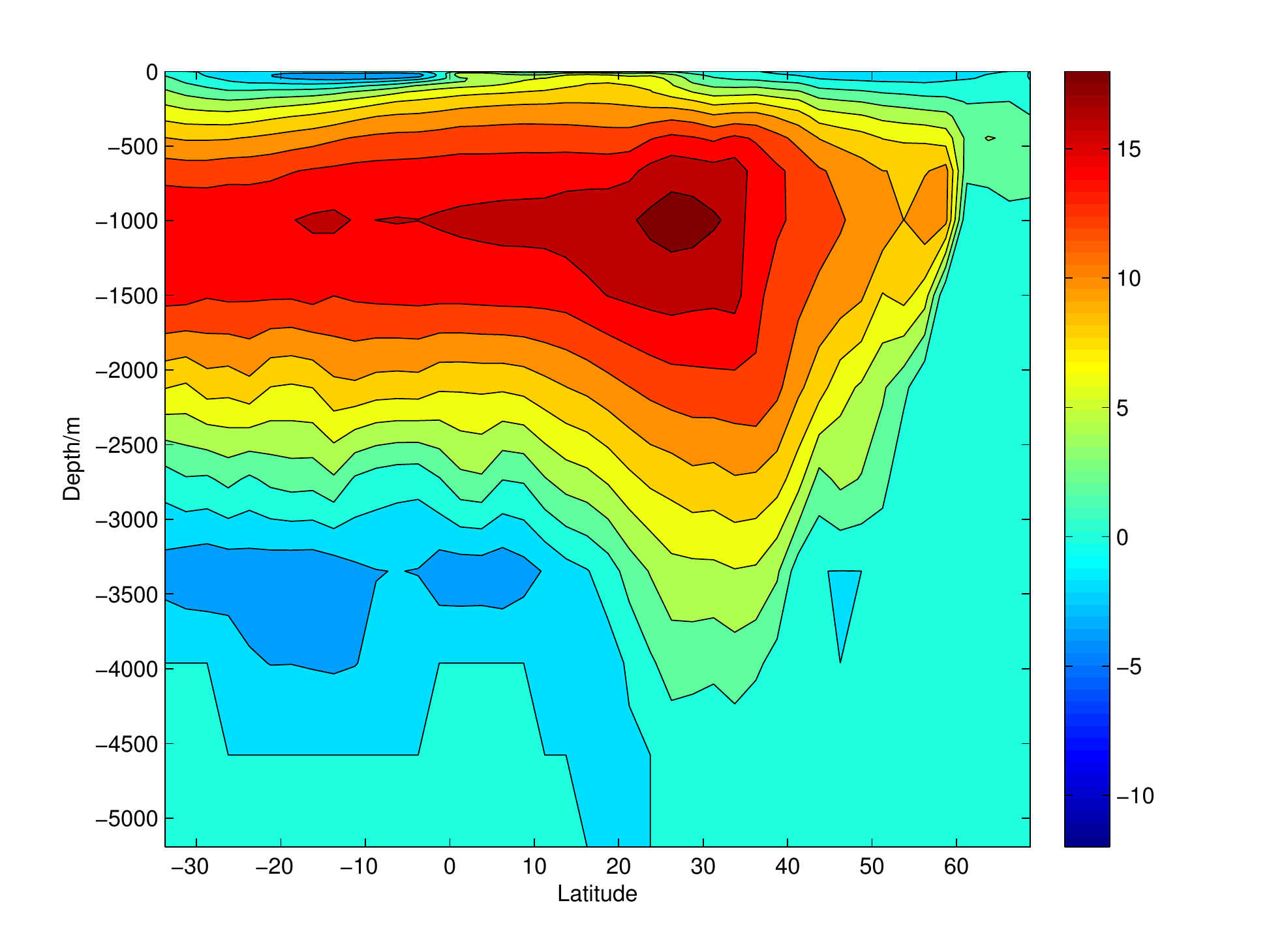} \textbf {(c)}  
        \includegraphics[width=0.4\textwidth]{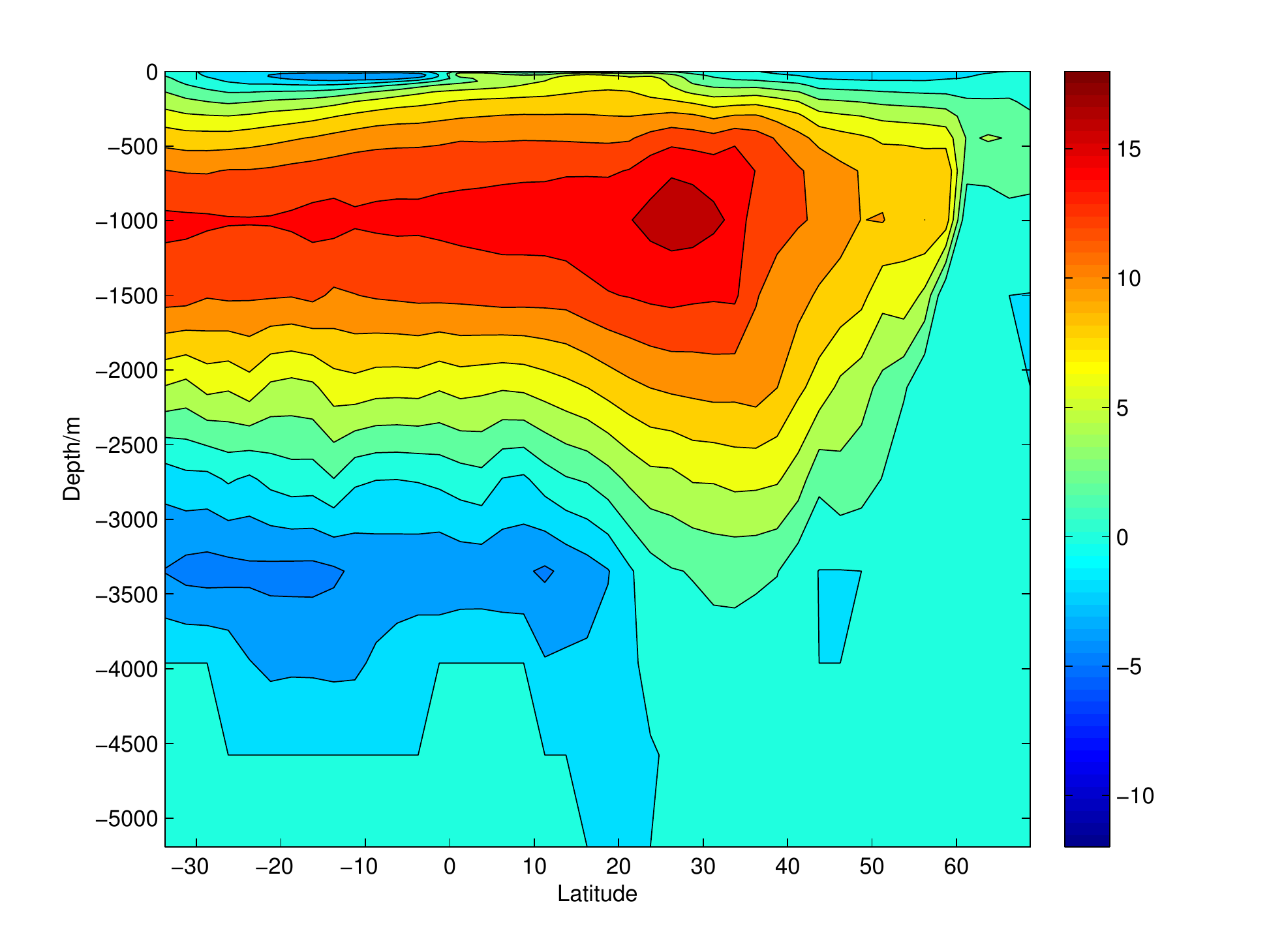} \textbf {(d)}
\end{center}
\caption{\it \small \textbf {(a)} Time series of the MOC (in Sv, 1 Sv = 10$^6$ m$^3$s$^{-1}$)  
at 26$^\circ$N and 1000m depth in the Atlantic  for the control simulation 
(green curve) and the freshwater perturbed simulation (blue curve) of the FAMOUS 
model. The red dots labeled from 1 to 6 show the average values of MOC from the 
corresponding equilibrium simulations and  the red broken line indicates the collapse
time $\tau_c = 874$ years.  \textbf {(b)} Annual mean MOC streamfunction pattern of equilibrium simulation 1. \textbf {(c)} Same 
as \textbf {(b)} but of equilibrium simulation 4.  \textbf {(d)} Same as \textbf {(b)} but of equilibrium simulation 6.  
}
\label{f:F1}
\end{figure}

\clearpage 

\begin{figure}[t]
\begin{center}
	\includegraphics[width=0.4\textwidth]{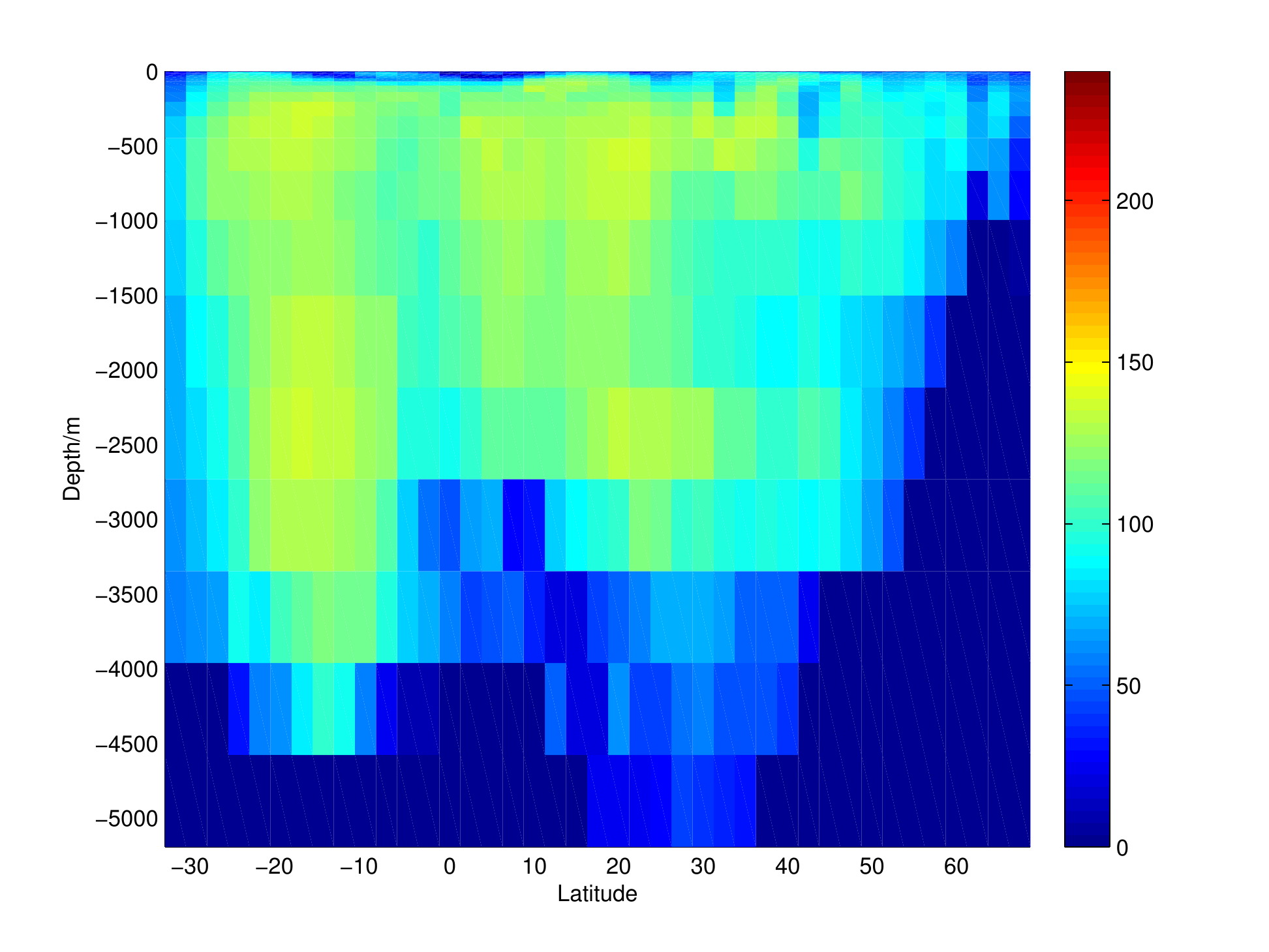} \textbf {(a)}
	\includegraphics[width=0.4\textwidth]{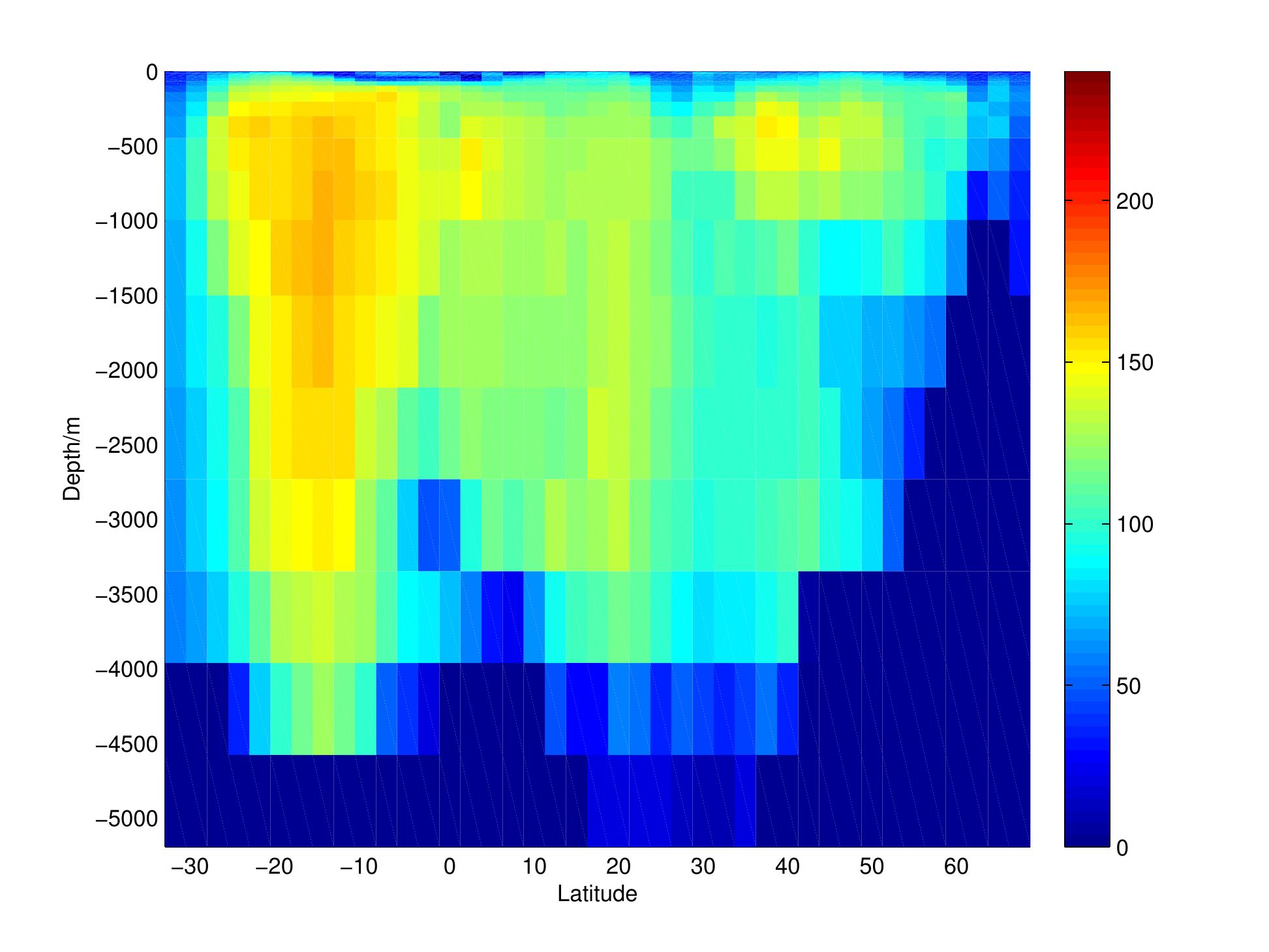}  \textbf {(b)} \\ 
	\includegraphics[width=0.4\textwidth]{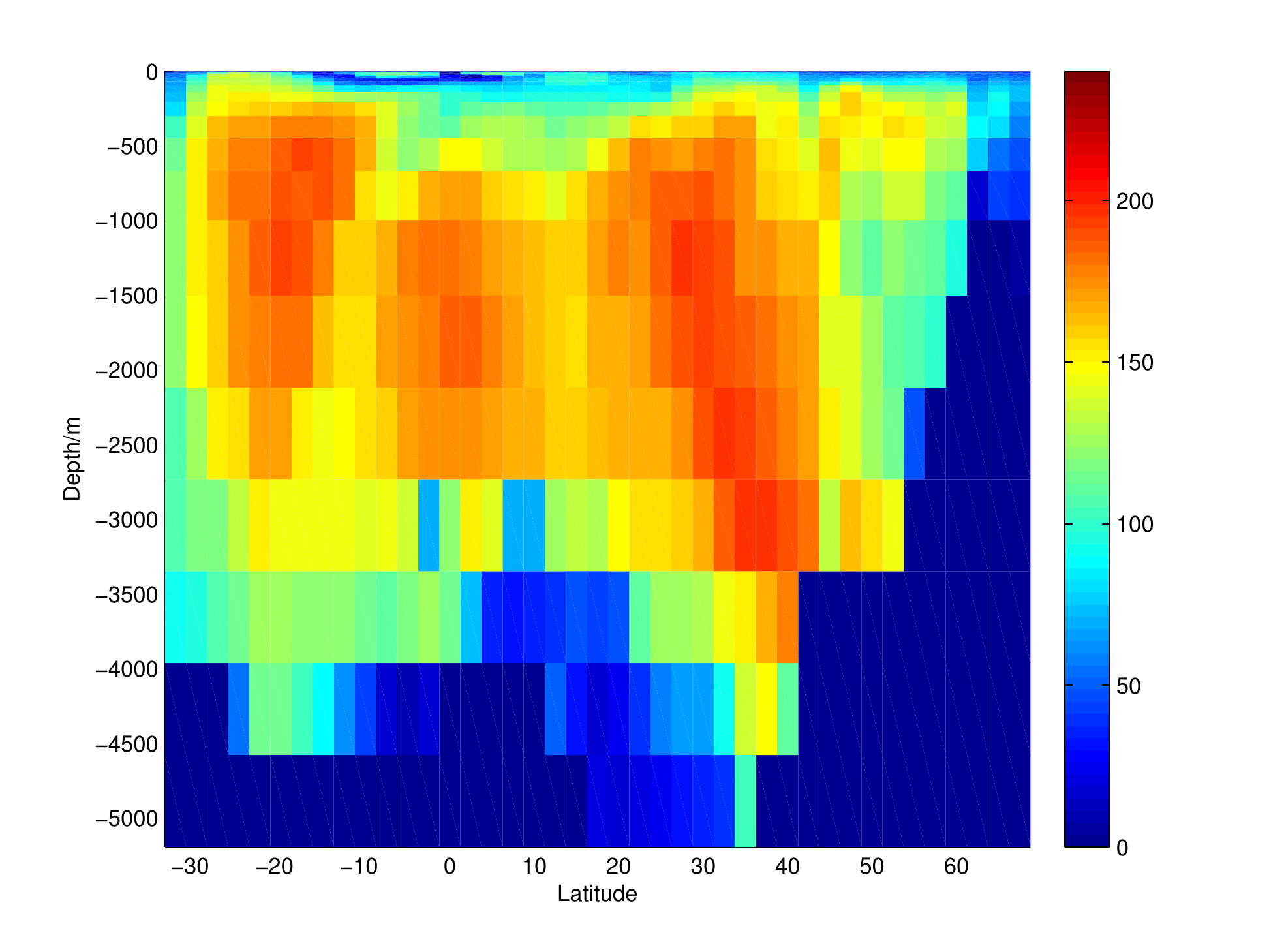} \textbf {(c)}
        \includegraphics[width=0.4\textwidth]{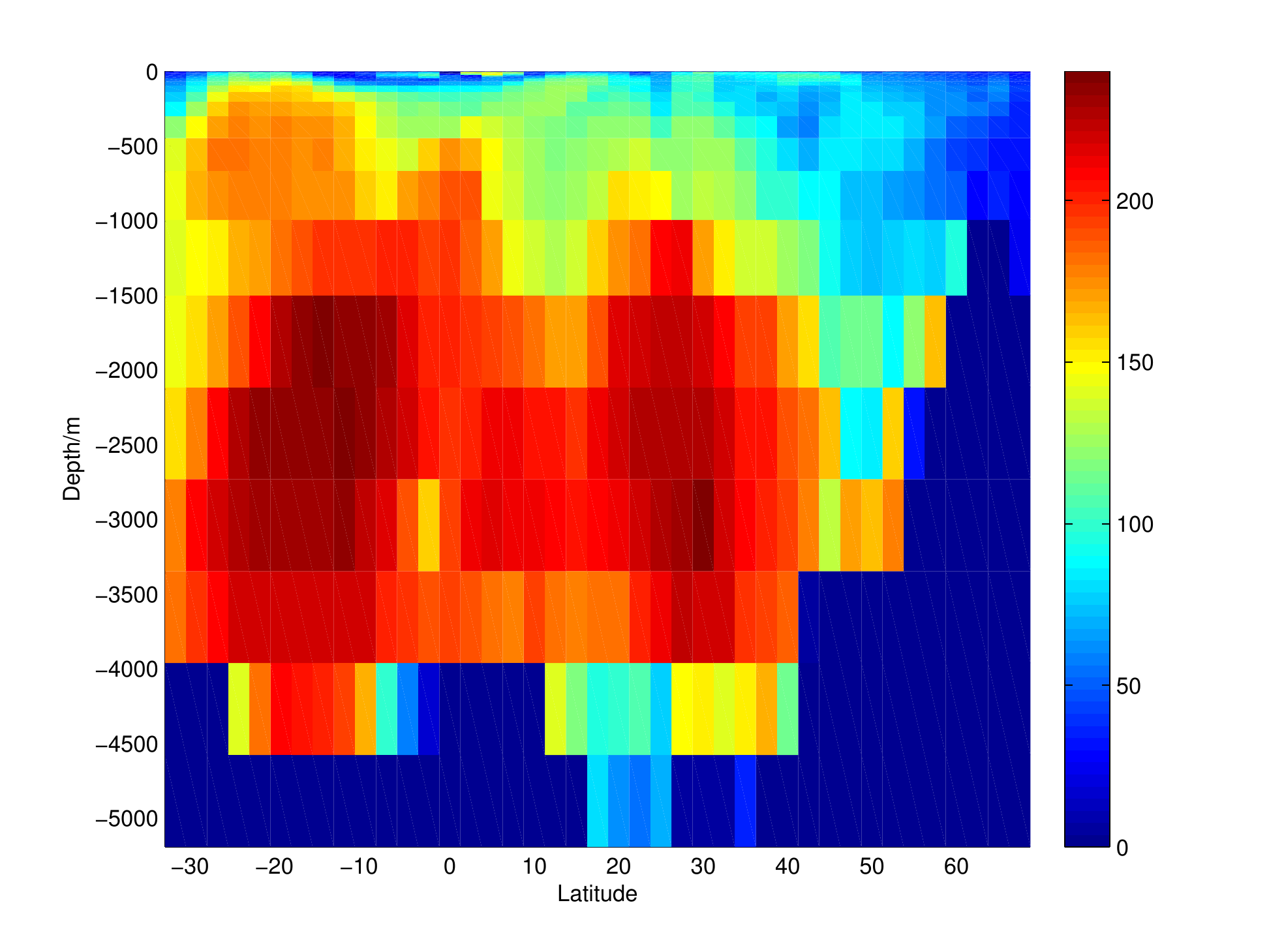} \textbf {(d)}
\end{center}
\caption{\it \small 
\textbf {(a)} Degree field of the Pearson Correlation Climate Network (PCCN) constructed 
from the MOC data of equilibrium simulation 1 in Fig.~1\textbf {a} using a threshold 
$\tau = 0.5$.   \textbf {(b)} Same as \textbf {(a)} but of equilibrium  simulation 2. \textbf {(c)} Same as \textbf {(a)} 
but of equilibrium simulation 4.  \textbf {(d)} Same as \textbf {(a)}  of equilibrium simulation 6.  }
\label{f:F2}
\end{figure}

\clearpage 
\begin{figure}[t]
\begin{center}
	\includegraphics[width=0.8\textwidth]{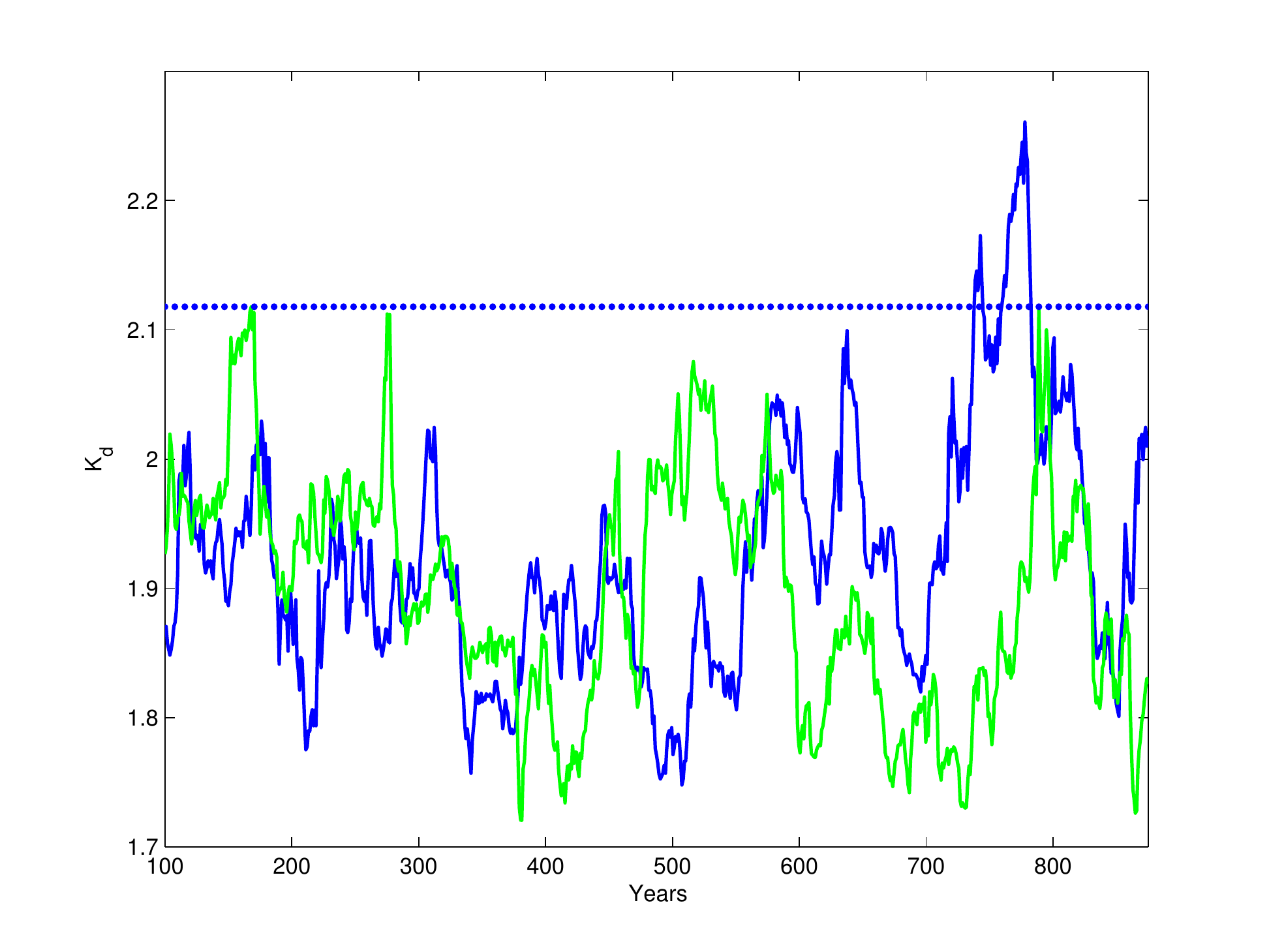} \textbf {(a)} \\
	\includegraphics[width=0.4\textwidth]{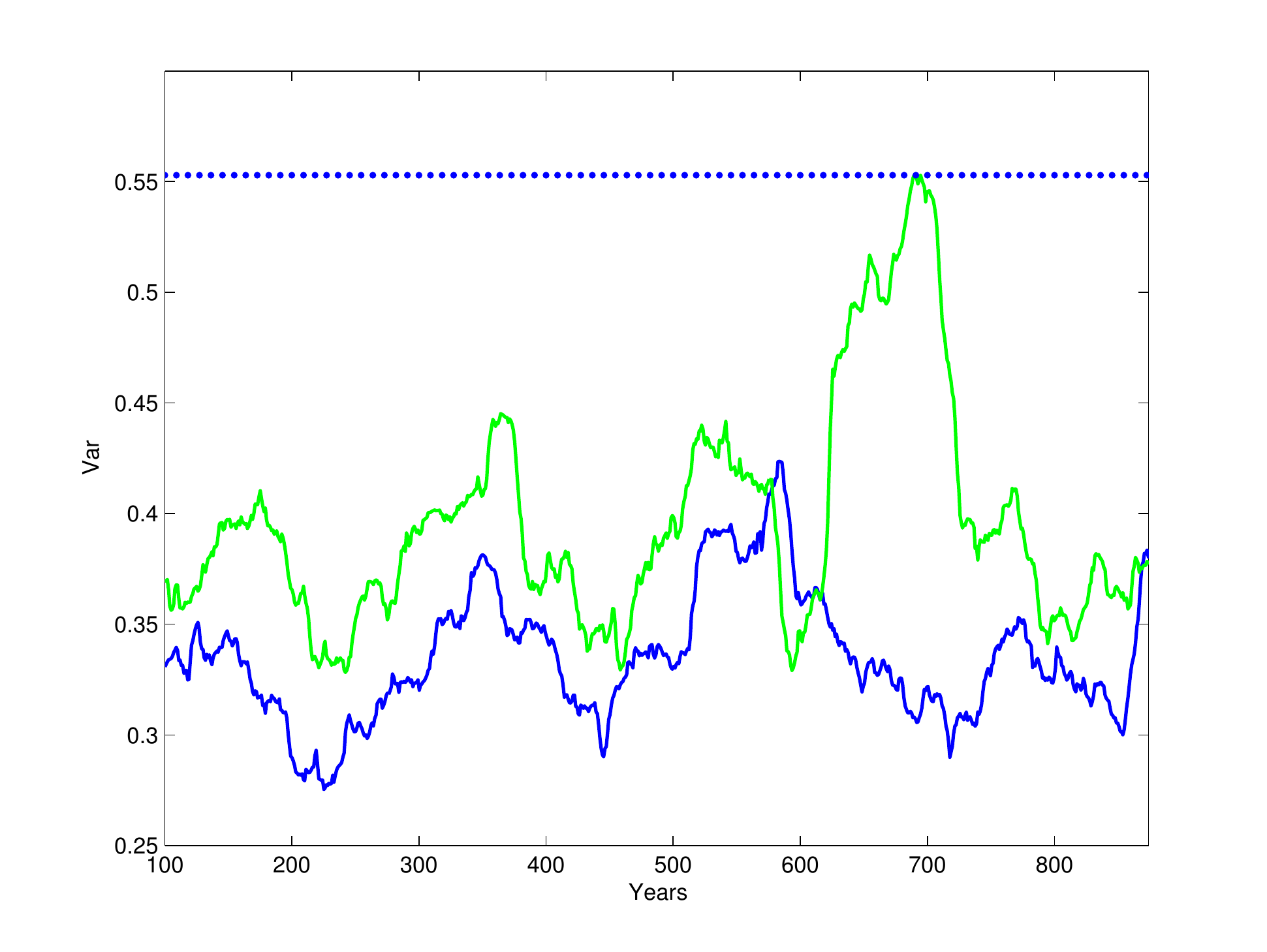} \textbf {(b)}
	\includegraphics[width=0.4\textwidth]{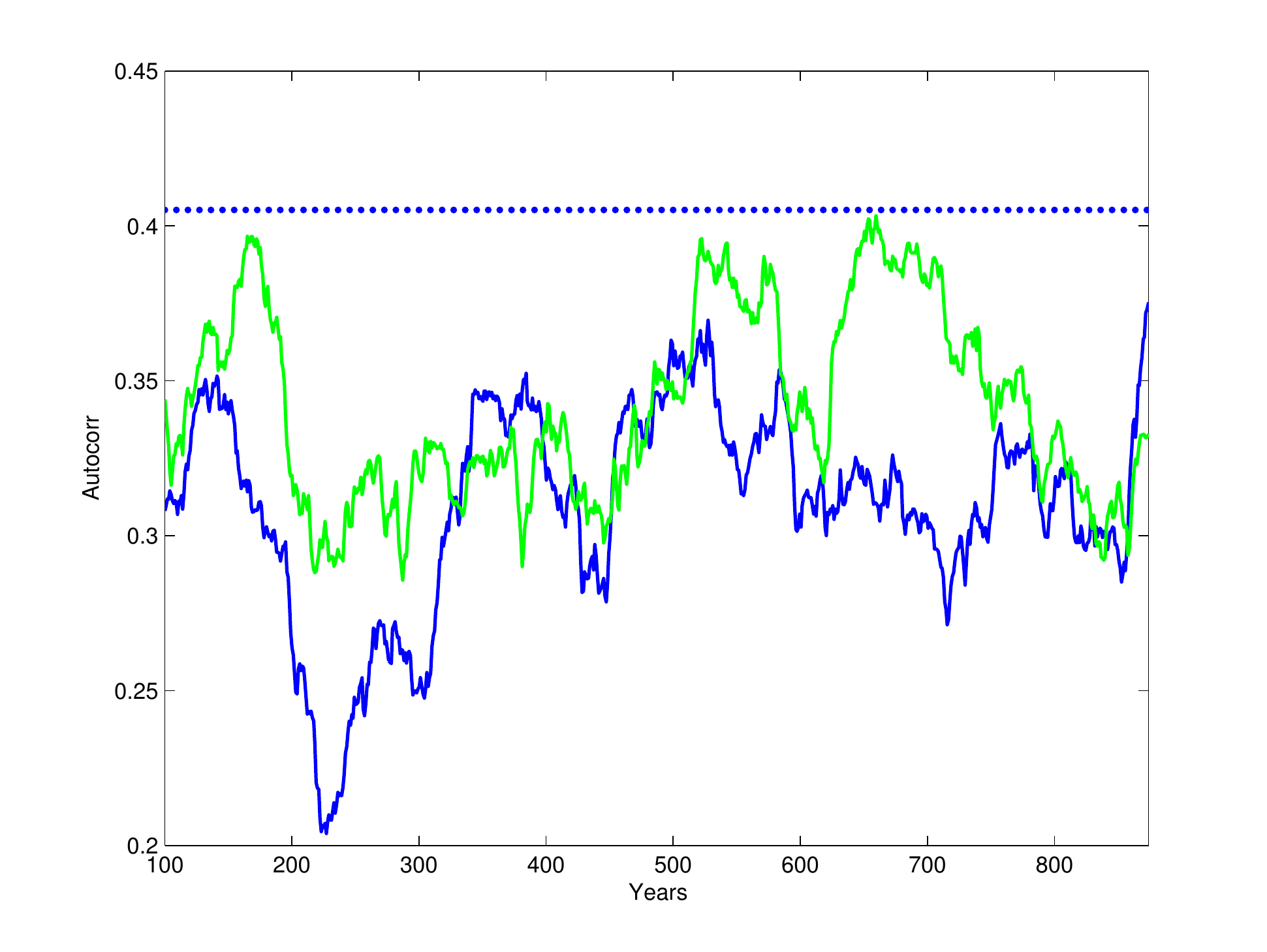} \textbf {(c)} \\
\end{center}
\caption{\it For the complete Atlantic MOC field,  \textbf {(a)} the kurtosis indicator 
$K_d$ gives the early warning signal at 738 years and lasts for 44 years. \textbf {(b)} The traditional 
variance indicator Var gives no early warning signal before the collapse time 
$\tau_c = 874$ years. \textbf {(c)} The traditional lag-1 autocorrelation indicator Autocorr gives 
no early warning signal before the collapse time $\tau_c$. Green solid curves 
are related to  the control simulation and blue solid curves are related to the  hosing 
simulation. The dashed blue horizontal lines indicate the corresponding maximum values of 
the control simulation.
  }
\label{f:F3}
\end{figure}

\clearpage 
\begin{figure}[t]
\begin{center}
	\includegraphics[width=1\textwidth]{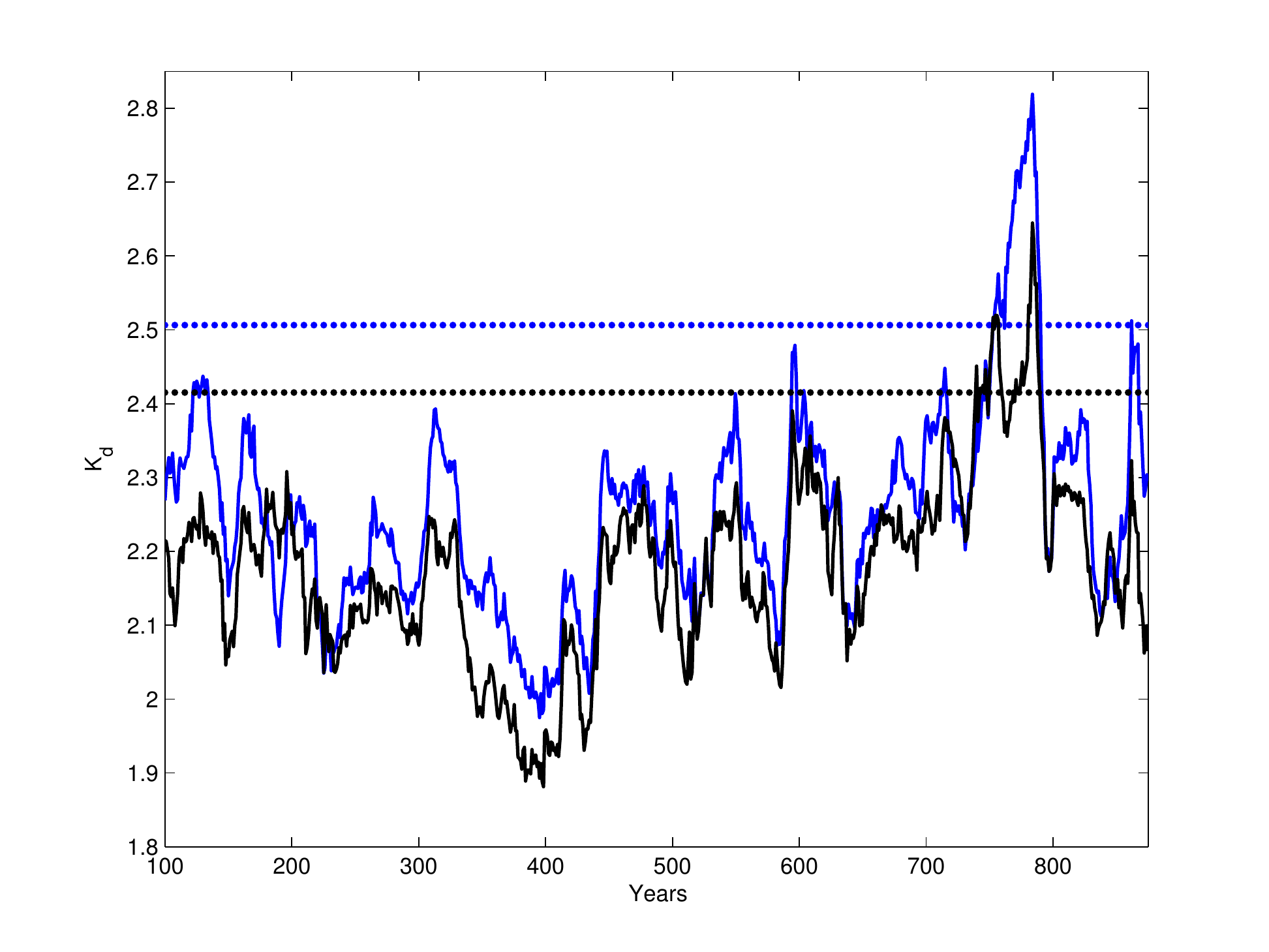} 
\end{center}
\caption{\it The kurtosis indicator $K_d$ for section data constructed from 
21 depth levels and 8 latitudes.  \textbf {(a)}  Data from 31$^\circ$N, 26$^\circ$N, 21$^\circ$N, 
16$^\circ$N, 11$^\circ$N, 8$^\circ$S, 13$^\circ$S and 18$^\circ$S (blue solid 
curve), gives the early warning signal at 753 years and lasts for 38 years.  
\textbf {(b)} Data from 31$^\circ$N, 26$^\circ$N, 21$^\circ$N, 16$^\circ$N, 11$^\circ$N, 
13$^\circ$S, 18$^\circ$S and 33$^\circ$S (black solid curve), 
gives the early warning signal at 740 years and lasts for 49 years.  The dotted blue 
horizontal line indicates the maximum $K_d$ value of the control simulation of  
section data \textbf {(a)},  and the dotted black horizontal line  indicates the 
corresponding maximum $K_d$ value  of section data \textbf {(b)}.  }
\label{f:F4}
\end{figure}

\clearpage 

\begin{table}[h]
\centering		
	\caption{\it The performance of the kurtosis indicator $K_d$ based on complex 
	networks  constructed from sets of section data at different latitudes. In the sets 
	of sections indicated by brackets, the first entry indicates the southern boundary, the 
	second entry the northern boundary and the third entry is the latitudinal step. }
 \begin{tabularx}{1\textwidth}{ X m{0.23\textwidth} m{0.19\textwidth} m{0.2\textwidth} m{0.2\textwidth}}
       		\hline
  			No. & Section & Detection Time (Year) & False Alarm & $\gamma^{K_d}$\\	 
  		\hline	
		    1  & (35$^\circ$S, 70$^\circ$N, 2.5$^\circ$)  &738 & No & 0.0675\\	
  			2 & 26$^\circ$N & 384/567/868 & Yes/Yes/No & 0.1031/0.0451/0.0105 \\ 	
  			3 & 33$^\circ$S & - & - & -0.1252 \\ 	
  			4 & (35$^\circ$S, 70$^\circ$N, 5$^\circ$) & 739 & No & 0.0472\\
  			5 & (20$^\circ$N, 70$^\circ$N, 5$^\circ$) & - & - & 0\\
  			6 & (35$^\circ$S, 20$^\circ$N, 5$^\circ$) & 785 & No & 0.0559\\
  			7 & (35$^\circ$S, 35$^\circ$N, 5$^\circ$) & 785 & No & 0.0773\\
  			8 & (20$^\circ$S, 35$^\circ$N, 5$^\circ$) & 785 & No & 0.0584\\
			9 &  Latitude set I & 753 & No & 0.1248\\
  			10 &  Latitudes set II & 740 & No & 0.0951\\
  			11 & Latitude set III & - & - & -0.0810 \\ 
  			12 & Latitude set IV & - & - &  -0.0454\\ 		
		\hline
	\end{tabularx}
	\begin{enumerate}[I]
		\item: 18$^\circ$S, 13$^\circ$S, 8$^\circ$S, 11$^\circ$N, 16$^\circ$N, 21$^\circ$N, 26$^\circ$N and 31$^\circ$N
		\item: 33$^\circ$S, 18$^\circ$S, 13$^\circ$S, 11$^\circ$N, 16$^\circ$N, 21$^\circ$N, 26$^\circ$N and 31$^\circ$N
		\item: 13$^\circ$N, 16$^\circ$N, 18$^\circ$NS, 21$^\circ$N, 23$^\circ$N, 26$^\circ$N, 28$^\circ$N and 31$^\circ$N
		\item: 33$^\circ$S, 31$^\circ$S, 28$^\circ$S, 26$^\circ$S, 23$^\circ$S, 21$^\circ$S, 18$^\circ$S and 16$^\circ$S			
	\end{enumerate}	 
\label{t:T1}
\end{table}


\clearpage 

\begin{thebibliography}{10}
\expandafter\ifx\csname url\endcsname\relax
  \def\url#1{\texttt{#1}}\fi
\expandafter\ifx\csname urlprefix\endcsname\relax\def\urlprefix{URL }\fi
\providecommand{\bibinfo}[2]{#2}
\providecommand{\eprint}[2][]{\url{#2}}

\bibitem{Ganachaud2000}
\bibinfo{author}{Ganachaud, A.} \& \bibinfo{author}{Wunsch, C.}
\newblock \bibinfo{title}{{Improved estimates of global ocean circulation, heat
  transport and mixing from hydrographic data}}.
\newblock \emph{\bibinfo{journal}{Nature}} \textbf{\bibinfo{volume}{408}},
  \bibinfo{pages}{453--457} (\bibinfo{year}{2000}).

\bibitem{Johns2011}
\bibinfo{author}{Johns, W.}, \bibinfo{author}{Baringer, M.} \&
  \bibinfo{author}{Beal, L.}
\newblock \bibinfo{title}{{Continuous, array-based estimates of Atlantic Ocean
  heat transport at 26.5 N}}.
\newblock \emph{\bibinfo{journal}{Journal Of Climate}}
  \textbf{\bibinfo{volume}{24}}, \bibinfo{pages}{2429--2449}
  (\bibinfo{year}{2011}).

\bibitem{Bryan1986}
\bibinfo{author}{Bryan, F.~O.}
\newblock \bibinfo{title}{{High-latitude salinity effects and interhemispheric
  thermohaline circulations}}.
\newblock \emph{\bibinfo{journal}{Nature}} \textbf{\bibinfo{volume}{323}},
  \bibinfo{pages}{301--304} (\bibinfo{year}{1986}).

\bibitem{Rahmstorf2000}
\bibinfo{author}{Rahmstorf, S.}
\newblock \bibinfo{title}{{The thermohaline circulation: a system with
  dangerous thresholds?}}
\newblock \emph{\bibinfo{journal}{Climatic Change}}
  \textbf{\bibinfo{volume}{46}}, \bibinfo{pages}{247--256}
  (\bibinfo{year}{2000}).

\bibitem{Stommel1961}
\bibinfo{author}{Stommel, H.}
\newblock \bibinfo{title}{{Thermohaline convection with two stable regimes of
  flow}}.
\newblock \emph{\bibinfo{journal}{Tellus}} \textbf{\bibinfo{volume}{2}},
  \bibinfo{pages}{244--230} (\bibinfo{year}{1961}).

\bibitem{Walin1985}
\bibinfo{author}{Walin, G.}
\newblock \bibinfo{title}{{The thermohaline circulation and the control of ice
  ages}}.
\newblock \emph{\bibinfo{journal}{Paleogeogr. Paleoclim. Paleoecol.}}
  \textbf{\bibinfo{volume}{50}}, \bibinfo{pages}{323--332}
  (\bibinfo{year}{1985}).


\bibitem{Lenton2008}
\bibinfo{author}{Lenton, T.~M.} \emph{et~al.}
\newblock \bibinfo{title}{Tipping elements in the earth's climate system}.
\newblock \emph{\bibinfo{journal}{Proceedings of the National Academy of
  Sciences}} \textbf{\bibinfo{volume}{105}}, \bibinfo{pages}{1786--1793}
  (\bibinfo{year}{2008}).

\bibitem{Lenton2011}
\bibinfo{author}{Lenton, T.~M.}
\newblock \bibinfo{title}{{Early warning of climate tipping points}}.
\newblock \emph{\bibinfo{journal}{Nature Publishing Group}}
  \textbf{\bibinfo{volume}{1}}, \bibinfo{pages}{201--209}
  (\bibinfo{year}{2011}).

\bibitem{Tsonis:2006tk}
\bibinfo{author}{Tsonis, A.~A.} \& \bibinfo{author}{Swanson, K.~L.}
\newblock \bibinfo{title}{{What do networks have to do with climate?}}
\newblock \emph{\bibinfo{journal}{Bulletin Of The American Meteorological
  Society}} \textbf{\bibinfo{volume}{87}}, \bibinfo{pages}{585--595}
  (\bibinfo{year}{2006}).

\bibitem{Donges2009}
\bibinfo{author}{Donges, J.~F.}, \bibinfo{author}{Zou, Y.},
  \bibinfo{author}{Marwan, N.} \& \bibinfo{author}{Kurths, J.}
\newblock \bibinfo{title}{{Complex networks in climate dynamics}}.
\newblock \emph{\bibinfo{journal}{The European Physical Journal Special
  Topics}} \textbf{\bibinfo{volume}{174}}, \bibinfo{pages}{157--179}
  (\bibinfo{year}{2009}).

\bibitem{Hawkins2011}
\bibinfo{author}{Hawkins, E.} \emph{et~al.}
\newblock \bibinfo{title}{{Bistability of the Atlantic overturning circulation
  in a global climate model and links to ocean freshwater transport}}.
\newblock \emph{\bibinfo{journal}{Geophysical Research Letters}}
  \textbf{\bibinfo{volume}{38}}, \bibinfo{pages}{L10605}
  (\bibinfo{year}{2011}).

\bibitem{Dakos2011}
\bibinfo{author}{Dakos, V.}, \bibinfo{author}{K{\'e}fi, S.},
  \bibinfo{author}{Rietkerk, M.}, \bibinfo{author}{van Nes, E.~H.} \&
  \bibinfo{author}{Scheffer, M.}
\newblock \bibinfo{title}{Slowing down in spatially patterned ecosystems at the
  brink of collapse}.
\newblock \emph{\bibinfo{journal}{The American Naturalist}}
  \textbf{\bibinfo{volume}{177}}, \bibinfo{pages}{154--166}
  (\bibinfo{year}{2011}).

\bibitem{Scheffer2009}
\bibinfo{author}{Scheffer, M.}, \bibinfo{author}{Bascompte, J.},
  \bibinfo{author}{Brock, W.~A.} \& \bibinfo{author}{Brovkin, V.}
\newblock \bibinfo{title}{{Early-warning signals for critical transitions}}.
\newblock \emph{\bibinfo{journal}{Nature}} \textbf{\bibinfo{volume}{461}},
  \bibinfo{pages}{53--59} (\bibinfo{year}{2009}).

\bibitem{Livina2007}
\bibinfo{author}{Livina, V.~N.} \& \bibinfo{author}{Lenton, T.~M.}
\newblock \bibinfo{title}{{A modified method for detecting incipient
  bifurcations in a dynamical system}}.
\newblock \emph{\bibinfo{journal}{Geophysical Research Letters}}
  \textbf{\bibinfo{volume}{34}} (\bibinfo{year}{2007}).

\bibitem{Mheen2013}
\bibinfo{author}{Mheen, M.} \emph{et~al.}
\newblock \bibinfo{title}{{Interaction network based early warning indicators
  for the Atlantic MOC collapse}}.
\newblock \emph{\bibinfo{journal}{Geophysical Research Letters}}
  \textbf{\bibinfo{volume}{40}}, \bibinfo{pages}{2714--2719}
  (\bibinfo{year}{2013}).

\bibitem{Tsonis2004}
\bibinfo{author}{Tsonis, A.~A.} \& \bibinfo{author}{Roebber, P.~J.}
\newblock \bibinfo{title}{{The architecture of the climate network}}.
\newblock \emph{\bibinfo{journal}{Physica A: Statistical Mechanics and its
  Applications}} \textbf{\bibinfo{volume}{333}}, \bibinfo{pages}{497--504}
  (\bibinfo{year}{2004}).

\bibitem{Feng2014}
\bibinfo{author}{Feng, Q.~Y.} \& \bibinfo{author}{Dijkstra, H.}
\newblock \bibinfo{title}{{Are North Atlantic multidecadal SST anomalies
  westward propagating?}}
\newblock \emph{\bibinfo{journal}{Geophysical Research Letters}}
  \textbf{\bibinfo{volume}{41}}, \bibinfo{pages}{541--546}
  (\bibinfo{year}{2014}).

\bibitem{Viebahn2014}
\bibinfo{author}{Viebahn, J.} \& \bibinfo{author}{Dijkstra, H.~A.}
\newblock \bibinfo{title}{{Critical Transition Analysis of the Deterministic
  Wind-Driven Ocean Circulation --- A Flux-Based Network Approach}}.
\newblock \emph{\bibinfo{journal}{International Journal of Bifurcation and
  Chaos}} \textbf{\bibinfo{volume}{24}}, \bibinfo{pages}{1430007}
  (\bibinfo{year}{2014}).

\bibitem{Cunningham2007}
\bibinfo{author}{Cunningham, S.~A.} \emph{et~al.}
\newblock \bibinfo{title}{{Temporal Variability of the Atlantic Meridional
  Overturning Circulation at 26.5 N}}.
\newblock \emph{\bibinfo{journal}{Science}} \textbf{\bibinfo{volume}{317}},
  \bibinfo{pages}{935--938} (\bibinfo{year}{2007}).

\bibitem{IPCC2013}
\bibinfo{author}{Collins, M.} \emph{et~al.}
\newblock \bibinfo{title}{{Long-term Climate Change: Projections, Commitments
  and Irreversibility}}.
\newblock In \bibinfo{editor}{Stocker, T.} \emph{et~al.} (eds.)
  \emph{\bibinfo{booktitle}{Climate Change 2013: The Physical Science Basis.
  Contribution of Working Group I to the Fifth Assessment Report of the
  Intergovernmental Panel on Climate Change}}, \bibinfo{pages}{1029--1136}
  (\bibinfo{publisher}{Cambridge: University of Cambridge Press Syndicate.},
  \bibinfo{address}{UK}, \bibinfo{year}{2013}).

\bibitem{Huisman2010}
\bibinfo{author}{Huisman, S.~E.}, \bibinfo{author}{den Toom, M.},
  \bibinfo{author}{Dijkstra, H.~A.} \& \bibinfo{author}{Drijfhout, S.}
\newblock \bibinfo{title}{{An Indicator of the Multiple Equilibria Regime of
  the Atlantic Meridional Overturning Circulation}}.
\newblock \emph{\bibinfo{journal}{Journal Of Physical Oceanography}}
  \textbf{\bibinfo{volume}{40}}, \bibinfo{pages}{551--567}
  (\bibinfo{year}{2010}).

\bibitem{Smeed2013}
\bibinfo{author}{Smeed, D.~A.} \emph{et~al.}
\newblock \bibinfo{title}{Observed decline of the atlantic meridional
  overturning circulation 2004 to 2012}.
\newblock \emph{\bibinfo{journal}{Ocean Science Discussions}}
  \textbf{\bibinfo{volume}{10}}, \bibinfo{pages}{1619--1645}
  (\bibinfo{year}{2013}).
\newblock \urlprefix\url{http://www.ocean-sci-discuss.net/10/1619/2013/}.

\bibitem{Kerr2000}
\bibinfo{author}{Kerr, R.~A.}
\newblock \bibinfo{title}{{A North Atlantic climate pacemaker for the
  centuries}}.
\newblock \emph{\bibinfo{journal}{Science}} \textbf{\bibinfo{volume}{288}},
  \bibinfo{pages}{1984--1986} (\bibinfo{year}{2000}).

\bibitem{Knight2005}
\bibinfo{author}{Knight, J.~R.}, \bibinfo{author}{Allan, R.~J.},
  \bibinfo{author}{Folland, C.~K.}, \bibinfo{author}{Vellinga, M.} \&
  \bibinfo{author}{Mann, M.~E.}
\newblock \bibinfo{title}{{A signature of persistent natural thermohaline
  circulation cycles in observed climate}}.
\newblock \emph{\bibinfo{journal}{Geophysical Research Letters}}
  \textbf{\bibinfo{volume}{32}}, \bibinfo{pages}{L20708}
  (\bibinfo{year}{2005}).

\end{thebibliography}
\end{document}


\noindent

\centerline{\Large Supplementary Material}  
\centerline{\Large to}  
\centerline{\Large Deep Ocean Early Warning Signals of } 
\centerline{\Large an Atlantic MOC Collapse} 
\bigskip
\centerline{\large by}  
\bigskip
\centerline{\large  Qing Yi Feng, Jan P. Viebahn and  Henk A. Dijkstra}

\vspace{1cm} 
This Supplementary Material contains the following:
\begin{enumerate}
\item [1.] Empirical Orthogonal Functions (EOFs) of the equilibrium MOC fields. 
\item [2.] Performance of $K_d$, variance and  autocorrelation indicators for 
single zonal sections. 
\item [3.] Sensitivity of the  $K_d$ indicator to different section data.   
\end{enumerate}

\clearpage

\section{The EOFs of the equilibrium MOC fields}

From the 100 year MOC data of the equilibrium simulations of the FAMOUS model 
at the labelled points  1 - 6   in Fig.~1\textbf {a}, the EOFs were computed using 
standard  methodology,  which we repeat here for convenience.  From the 
$n \times N$ (with $n = 100$ and $N = 21 \times 42 = 882$) data  matrix  $F$, 
ordered such that  each column contains  the MOC anomaly time series of 
a grid point,  we form the  $N \times N$  covariance matrix  $\Sigma$ by calculating  
\be
\Sigma = F^T F , 
\ee
where the superscript $T$ indicates the transpose. Next, we solve the 
eigenvalue problem 
\be
\Sigma {\bf e}_i =  \lambda_i {\bf e}_i , 
\ee
where the $\lambda_i$ are the eigenvalues of $\Sigma$, with 
$i = 1, \cdots, N$ and the  eigenvectors ${\bf e}_i$  are the  
EOFs. 

\begin{figure}[htbp]
\begin{center}
	\includegraphics[width=0.4\textwidth]{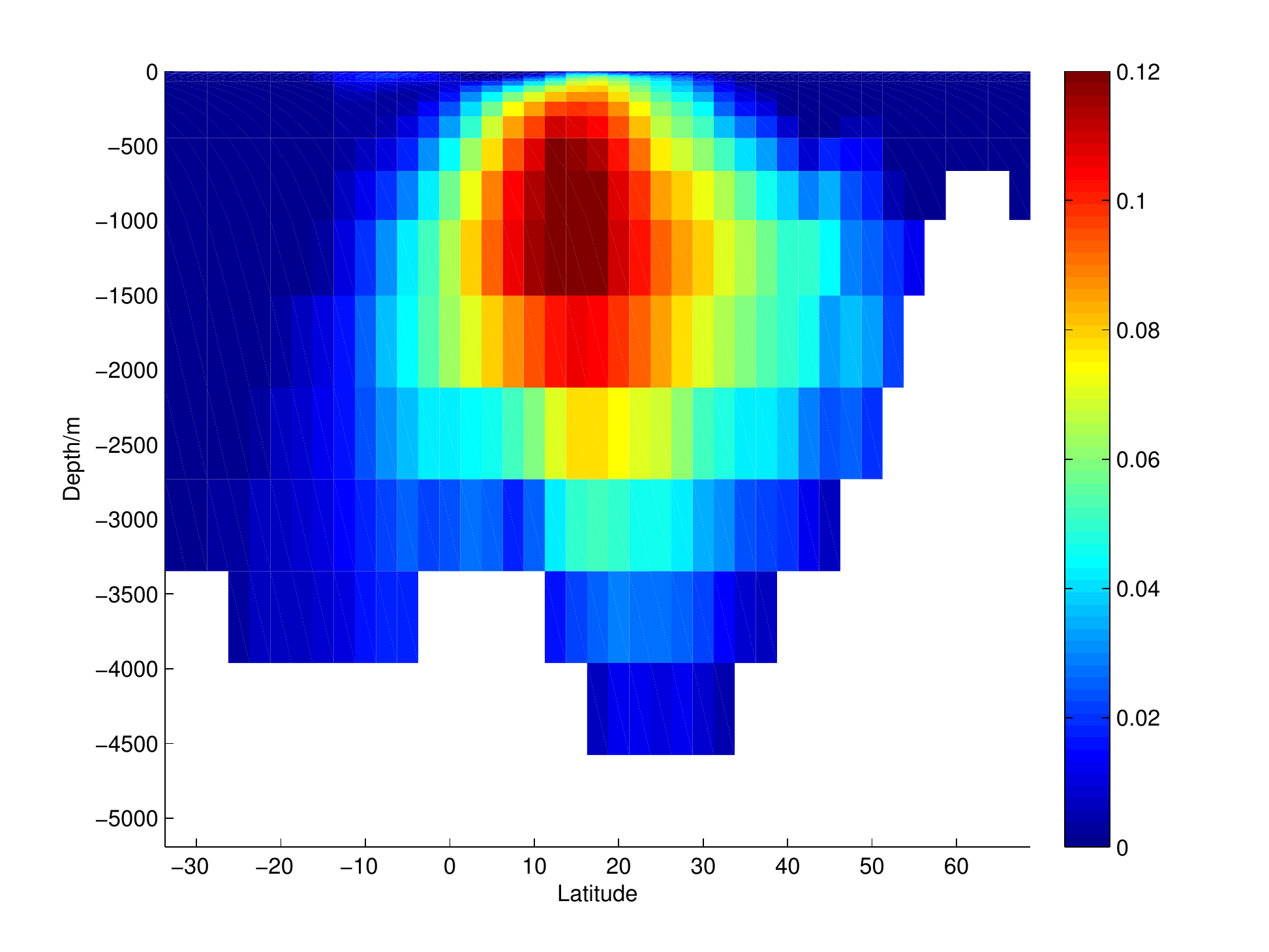} \textbf {(a)}
	\includegraphics[width=0.4\textwidth]{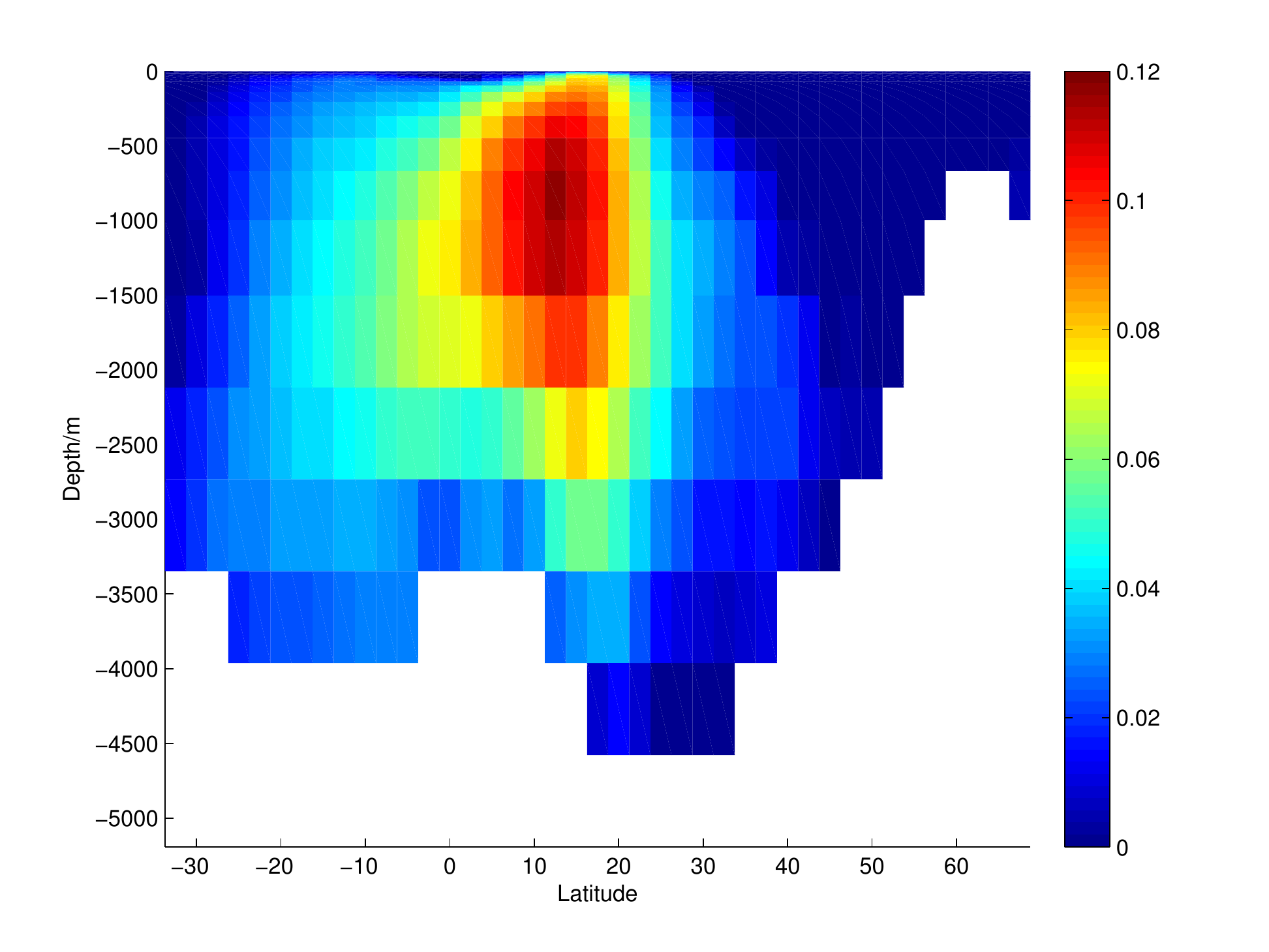} \textbf {(b)}\\
	\includegraphics[width=0.4\textwidth]{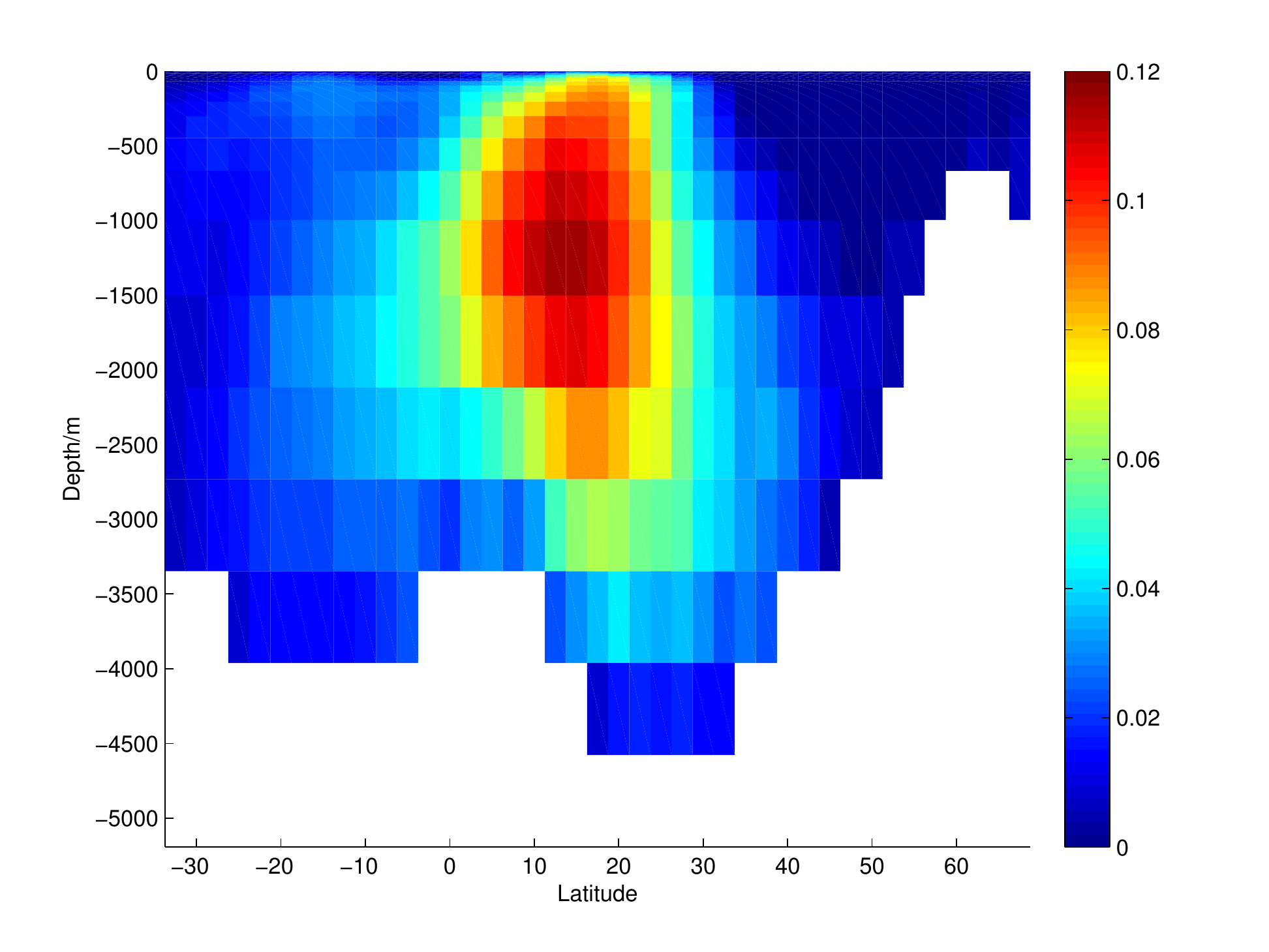} \textbf {(c)} 
	\includegraphics[width=0.4\textwidth]{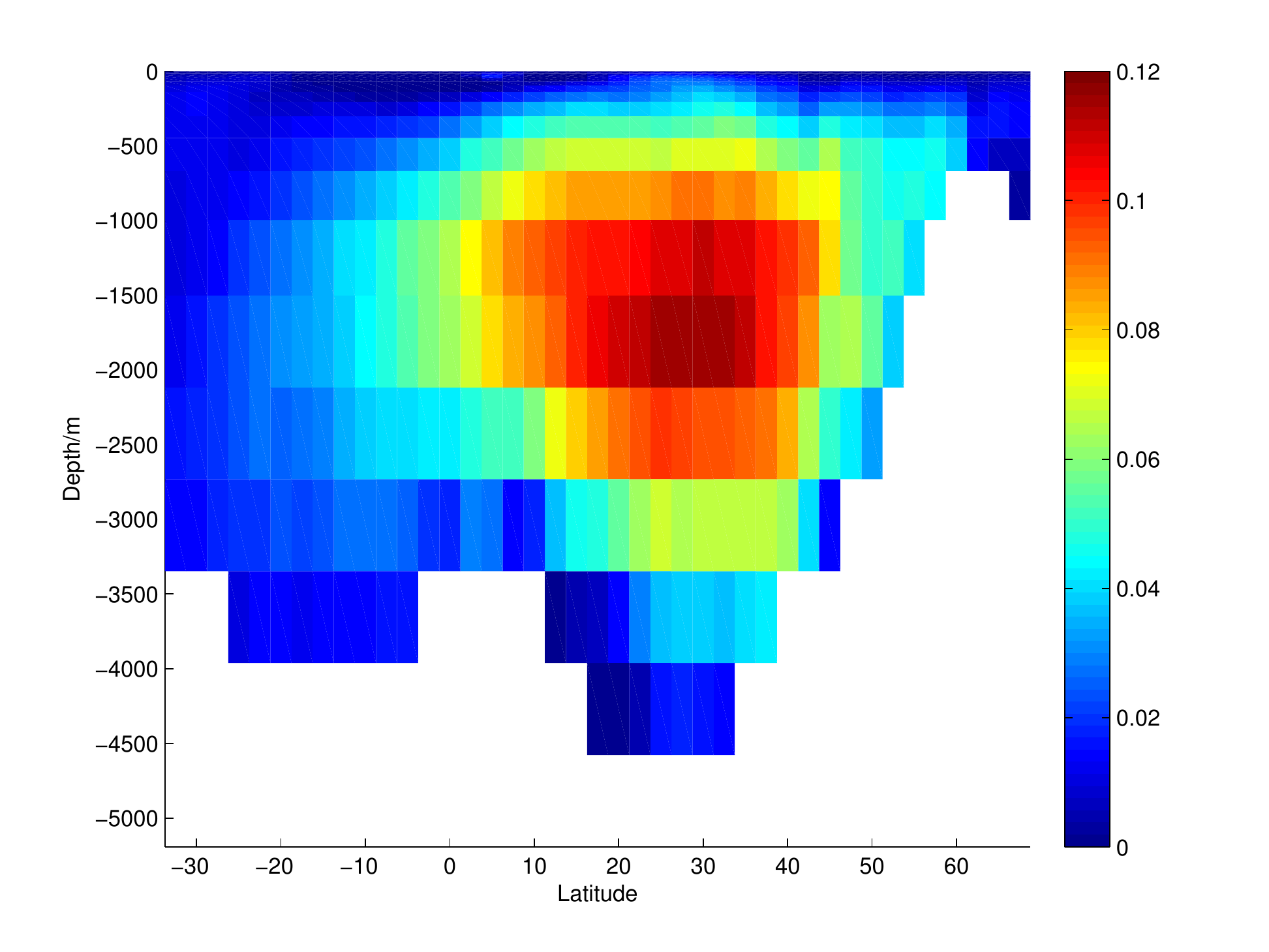} \textbf {(d)}\\
	\includegraphics[width=0.4\textwidth]{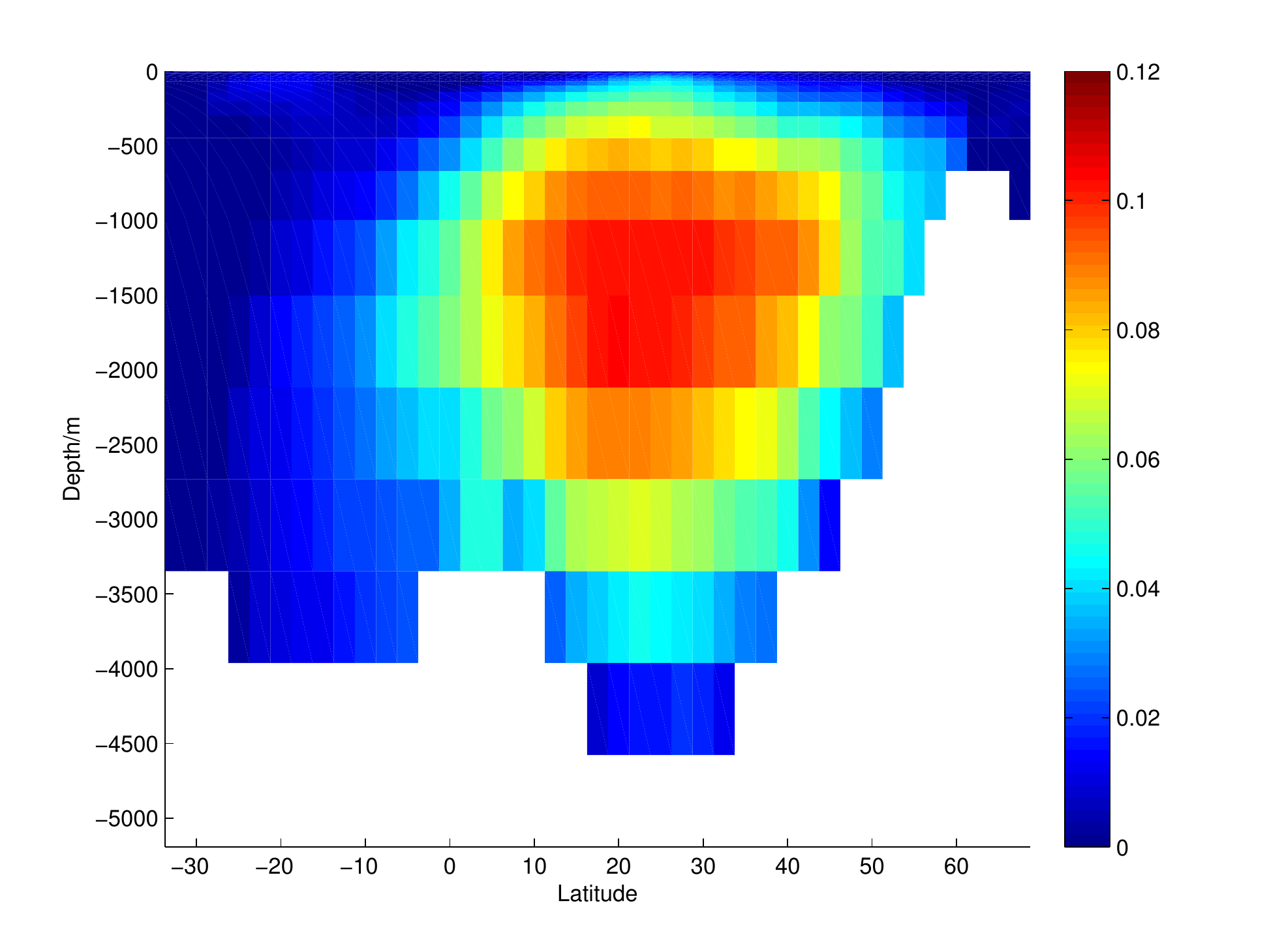} \textbf {(e)} 
	\includegraphics[width=0.4\textwidth]{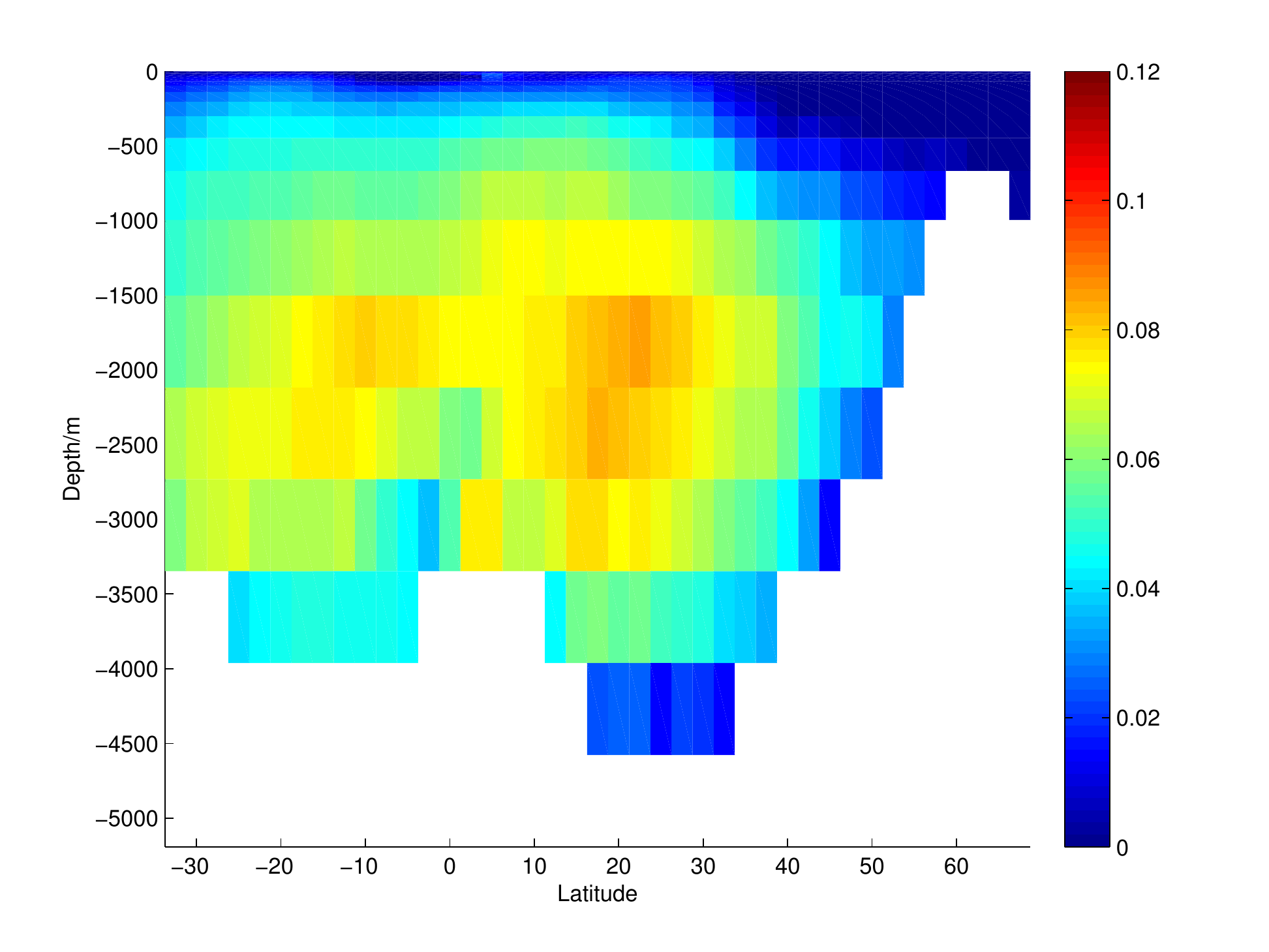} \textbf {(f)}\\
\end{center}
\caption{\it \small \textbf {(a)} The first EOF of the MOC data of equilibrium 
simulation 1, explaining 26.3\%  of the variance.  
\textbf {(b)} Same as \textbf {(a)} but of equilibrium simulation 2, explaining 24.1\%  of the variance. 
\textbf {(c)} Same as \textbf {(a)} but of equilibrium simulation 3, explaining 24.6\% of the variance. 
\textbf {(d)} Same as \textbf {(a)} but of equilibrium simulation 4, explaining 37.1\% of the variance. 
\textbf {(e)} Same as \textbf {(a)} but of equilibrium simulation 5, explaining 30.5\% of the variance. 
\textbf {(f)} Same as \textbf {(a)} but of equilibrium simulation 6, explaining 37.6\% of the variance. }
\label{f:EOFs}
\end{figure}

The amount of variance $\sigma_i$ explained by  each EOF  is obtained from 
\be
\sigma_i = \frac{\lambda_i}{\sum_{j = 1}^N  \lambda_j} \times 100\%
\ee
and the EOFs are typically ordered according to decreasing variance. 
Supplementary Fig.~\ref{f:EOFs} shows the first EOF of each of the six different 
equilibrium simulations  with the values of $\sigma_i$ shown in the 
caption.  

The first EOF becomes more dominant (i.e., the value of explained
variance increases) when the freshwater forcing is increased (from 
1 to 6).  From Supplementary 
Fig.~\ref{f:EOFs} one observes that for small freshwater forcing 
(panels \textbf {a}-\textbf {b}) most of the variability takes place near 15$^\circ$ 
and at about 1000 m depth. However, the maximum shifts to higher
latitudes and deeper locations when the freshwater in increased 
(panels \textbf {c}-\textbf {e}) and closest to the transition  (panel \textbf {f})
the maximum occurs at about 2000 m depth at both 20$^\circ$S 
and 20$^\circ$N. 

The principal component (PC) time  series of an EOF   
can be found  by calculating
\be
{\bf a}_i = F {\bf e}_i
\ee
such that  the data can  be reconstructed from the EOFs and PCs by
\be
F = \sum_{j = 1}^N {\bf a}_j {\bf e}^T_j
\ee
This reconstruction can be used to  determine how the different EOFs 
affect   the degree  distribution  the network. For example, one can consider 
\be
F_1 = {\bf a}_1 {\bf e}^T_1 ~ ; ~ F_2 = {\bf a}_1 {\bf e}^T_1 + {\bf a}_2 {\bf e}^T_2 ~ ; ~ 
F_3 = {\bf a}_1 {\bf e}^T_1 + {\bf a}_2 {\bf e}^T_2 + {\bf a}_3 {\bf e}^T_3
\ee
and reconstruct  networks from each dataset $F_i$ using the same 
methods described in the main text. 

As all correlation coefficients   for the dataset $F_1$ are 
unity,  an  interaction network based on  $F_1$ is fully connected.  
The networks for $F_2$ and $F_3$  are  not fully connected which 
indicates that as more EOFs explain the  variance in the time series, 
the degree distribution deviates further from  that of the maximum 
degree.  This illustrates that when the tipping  point is approached 
and one EOF tends to dominate in the time series,  there will be 
regions of  very high degree. 

\section{Performance of $K_d$, variance and  autocorrelation indicators for 
single sections.}

In Supplementary Table~\ref{t:T1}, we show the performance of the kurtosis 
indicator $K_d$  constructed from 21 different single zonal sections.  We 
excluded single sections north of 55$^\circ$N as the MOC amplitude is very 
small here. 
\begin{table}[htbp]
\centering		
	\caption{\it The performance of the kurtosis indicator $K_d$ based on complex networks 
	reconstructed from different single zonal sections}
\begin{tabularx}{1\textwidth}{ X m{0.09\textwidth} m{0.08\textwidth} m{0.09\textwidth} | m{0.09\textwidth} m{0.12\textwidth} m{0.12\textwidth} m{0.12\textwidth}}
       		\hline
  			Section & Detection Time (Year) & False Alarm & $\gamma^{K_d}$ & Section & Detection Time (Year) & False Alarm & $\gamma^{K_d}$\\	 
		\hline
   			33$^\circ$S & - & - & -0.1252 & 11$^\circ$N & - & - & -0.0202\\
    			28$^\circ$S & - & - & -0.1540 & 16$^\circ$N & 872 & No & 0.1792\\ 
   			23$^\circ$S & - & - & -0.0090 & 21$^\circ$N & 396/866 & Yes/No & 0.0045/0.0420\\ 
   			18$^\circ$S & - & - & -0.0456 & 26$^\circ$N & 384/567/868 & Yes/Yes/No & 0.1031/0.0451/0.0105\\ 
   			13$^\circ$S & - & - & -0.1774 & 31$^\circ$N & 384 & Yes & 0.0161\\ 
   			8$^\circ$S & - & - & -0.0595 & 36$^\circ$N & 382 & Yes & 0.0909\\ 
   			3$^\circ$S & - & - & -0.1937 & 41$^\circ$N & 586$^{\textbf {(a)}}$ & Yes & 0.0057\\ 
   			1$^\circ$N & - & - & 0 & 46$^\circ$N & - & - & -0.1463\\ 
   			6$^\circ$N & - & - & -0.0840 & 51$^\circ$N & - & - & -0.0260\\ 
  	\hline
	\end{tabularx}
		\begin{enumerate}[(a)]
		\item Signal only lasts for 2 years.	
	\end{enumerate}	 
\label{t:T1}
\end{table}

The performance of the variance indicator (Var) is shown in 
Supplementary Table~\ref{t:T2}. 
Obviously, the variance is not able to detect the MOC collapse because  $\gamma^{Var}$
is negative for each zonal section.  The lag-1 autocorrelation (AC), shown in 
Supplementary Table~\ref{t:T3},  provides an  early warning signal at a few 
zonal sections. However, the detection is   either too early (false alarm) as for 
21$^\circ$N and  23$^\circ$S or very  late (as for 13$^\circ$S).  
\begin{table}[htbp]
\centering		
	\caption{\it The performance of the variance indicator Var at different 
	                  single zonal sections}
\begin{tabularx}{1\textwidth}{ X m{0.1\textwidth} m{0.1\textwidth} m{0.1\textwidth} | m{0.1\textwidth} m{0.1\textwidth} m{0.1\textwidth} m{0.1\textwidth}}
       		\hline
  			Section & Detection Time (Year) & False Alarm & $\gamma^{Var}$ & Section & Detection Time (Year) & False Alarm & $\gamma^{Var}$\\	 
		\hline
   			33$^\circ$S & - & - & -0.0694 &11$^\circ$N & - & - & -0.1505\\ 
    			28$^\circ$S & - & - & -0.0697 & 16$^\circ$N & - & - & -0.2153\\ 
   			23$^\circ$S & - & - & -0.0747 & 21$^\circ$N & - & - & -0.3104\\ 
   			18$^\circ$S & - & - & -0.0868 & 26$^\circ$N & - & - & -0.3773\\ 
   			13$^\circ$S & - & - & -0.0500 & 31$^\circ$N & - & - & -0.3389\\ 
   			8$^\circ$S & - & - & -0.0515 & 36$^\circ$N & - & - & -0.3116\\ 
   			3$^\circ$S & - & - & -0.0326 & 41$^\circ$N & - & - & -0.3482\\ 
   			1$^\circ$N & - & - & -0.0402 & 46$^\circ$N & - & - & -0.4464\\ 
   			6$^\circ$N & - & - & -0.1279 & 51$^\circ$N & - & - & -0.4735\\ 

  	\hline
	\end{tabularx}
\label{t:T2}
\end{table}
\begin{table}[htbp]
\centering		
	\caption{\it The performance of the Lag-1 Autocorrelation indicator AC  at 
	                   different single zonal sections}
\begin{tabularx}{1\textwidth}{ X m{0.1\textwidth} m{0.1\textwidth} m{0.1\textwidth} | m{0.1\textwidth} m{0.1\textwidth} m{0.1\textwidth} m{0.1\textwidth}}
       		\hline
  			Section & Detection Time (Year) & False Alarm & $\gamma^{AC}$ & Section & Detection Time (Year) & False Alarm & $\gamma^{AC}$\\	 
		\hline
   			33$^\circ$S & - & - & -0.1470 &11$^\circ$N & - & - & -0.1168\\ 
    			28$^\circ$S & - & - & -0.0265 & 16$^\circ$N & - & - & -0.0512\\ 
   			23$^\circ$S & 497 & Yes & 0.0593 & 21$^\circ$N & 506 & Yes & 0.1769\\ 
   			18$^\circ$S & - & - & -0.0987 & 26$^\circ$N & - & - & -0.0471\\ 
   			13$^\circ$S & 870 & No & 0.0416 & 31$^\circ$N & - & - & -0.0654\\ 
   			8$^\circ$S & - & - & -0.0582 & 36$^\circ$N & - & - & -0.0820\\ 
   			3$^\circ$S & - & - & -0.0904 & 41$^\circ$N & - & - & -0.0455\\ 
   			1$^\circ$N & - & - & -0.0993 & 46$^\circ$N & - & - & -0.1420\\ 
   			6$^\circ$N & - & - & -0.0068 & 51$^\circ$N & - & - & -0.1107\\ 
  	\hline
	\end{tabularx}
\label{t:T3}
\end{table}

\section{Sensitivity of the  $K_d$ indicator to different section data. }

To further investigate the contribution of different sections to the 
detection of the MOC collapse, we exclude single sections one by one from 
the optimal locations we found (the set of eight sections labelled I in Table~1 
in the main text),  and evaluate the performance of the  indicator $K_d$ for 
the network  reconstructed from the remaining seven locations. 

As shown in Supplementary Table~\ref{t:T4}, sections 18$^\circ$S and 
31$^\circ$N are essential for any  detection. The data at sections 13$^\circ$S and 
8$^\circ$S  enhance the performance of the indicator $K_d$, 
by increasing the value of $\gamma^{K_d}$. The data at  sections 
11$^\circ$N and 16$^\circ$N  prevent  the occurrence  of false 
alarms and the data at  sections 21$^\circ$N and 26$^\circ$N  decrease
the detection time.
\begin{table}[htbp]
\centering		
	\caption{\it Sensitivity test of the kurtosis indicator $K_d$ by excluding, 
	in each case, one single section from the set I of optimal sections given in 
   Table 1 in the main text.}
 \begin{tabularx}{1\textwidth}{ X m{0.21\textwidth} m{0.19\textwidth} m{0.2\textwidth} }
       		\hline
  			Section & Detection Time (Year) & False Alarm & $\gamma^{K_d}$\\	 
  		\hline
   			7 latitudes (I$-$18$^\circ$S) & - & - & -0.0303\\ 
  			7 latitudes (I$-$13$^\circ$S) & 768 & No & 0.0233\\ 
  			7 latitudes (I$-$8$^\circ$S) & 773 & No & 0.0047\\ 
  			7 latitudes (I$-$11$^\circ$N) & 596$^{\textbf {(a)}}$/712$^{\textbf {(b)}}$/772$^{\textbf {(c)}}$ & Yes/Yes/Yes & 0.0175/0.0096/0.0163\\ 
  			7 latitudes (I$-$16$^\circ$N) & 443 & Yes & 0.1303\\ 
  			7 latitudes (I$-$21$^\circ$N) & 862 & No & 0.0977\\ 
  			7 latitudes (I$-$26$^\circ$N) & 859 & No & 0.0317\\ 
  			7 latitudes (I$-$31$^\circ$N) & - & - & -0.0033\\
  	\hline
	\end{tabularx}
	\begin{enumerate}[(a)]
		\item Signal only lasts for 2 years.
		\item Signal only lasts for 1 years.
		\item Signal only lasts for 3 years.
	\end{enumerate}	 
\label{t:T4}
\end{table}